%% file: vqsd.tex
\documentclass[prx,twocolumn,floatfix,superscriptaddress,longbibliography]{revtex4-1}

\usepackage{amssymb,amsmath,amstext}
\usepackage{graphicx}
\usepackage{epstopdf}
\usepackage{color}
\usepackage{appendix}
\usepackage[T1]{fontenc}
\usepackage{bbold}
\usepackage{bbm}
\usepackage{latexsym}
\usepackage[colorlinks=true,citecolor=blue,linkcolor=magenta]{hyperref}
\usepackage[latin1]{inputenc}
\usepackage{tikz}

\usetikzlibrary{shapes,arrows}
\renewcommand{\vec}[1]{\boldsymbol{#1}}  
\newcommand{\<}{\langle}
\renewcommand{\>}{\rangle}

\newcommand{\U}{U_{p}(\vec{\alpha})}
\newcommand{\Udag}{U^\dagger_{p}(\vec{\alpha})}
\newcommand{\Uopt}{U_{p}(\vec{\alpha}_{\text{opt}})}

\long\def\ca#1\cb{} 

\newcommand{\avg}[1]{\langle #1\rangle }
\newcommand{\ket}[1]{|#1\rangle}               
\newcommand{\bra}[1]{\langle #1|}              
\newcommand{\dya}[1]{\ket{#1}\!\bra{#1}}
\newcommand{\dyad}[2]{\ket{#1}\!\bra{#2}}        
\newcommand{\ip}[2]{\langle #1|#2\rangle}      

\newcommand{\DC}{\mathcal{D}}

\newcommand{\SC}{\mathcal{S}}

\newcommand{\ZC}{\mathcal{Z}}

\newcommand{\HS}{\text{HS}}

\newcommand{\Tr}{{\rm Tr}}

\renewcommand{\geq}{\geqslant}
\renewcommand{\leq}{\leqslant}
\newcommand{\mte}[2]{\langle#1|#2|#1\rangle }
\newcommand{\mted}[3]{\langle#1|#2|#3\rangle }

\DeclareMathOperator*{\argmin}{arg\,min}
\renewcommand{\vec}[1]{\boldsymbol{#1}}  
\newcommand{\ot}{\otimes}
\newcommand{\ad}{^\dagger}

\newcommand*{\id}{\openone}

\newcommand{\rhot}{\tilde{\rho}}

\begin{document}

\usetikzlibrary{calc}

\tikzstyle{decision} = [diamond, draw, fill=purple!60, 
    text width=4.5em, text badly centered, node distance=3cm, inner sep=0pt, rounded corners, minimum height=7em, minimum width = 7em]
\tikzstyle{block} = [rectangle, draw, fill=blue!60, 
    text width=5em, text centered, rounded corners, minimum height=4em]
\tikzstyle{block2} = [rectangle, draw, fill=orange!60, 
    text width=5em, text centered, rounded corners, minimum height=4em]
\tikzstyle{line} = [draw, -latex']
\tikzstyle{cloud} = [draw, ellipse, fill=pink!150, node distance=3cm,
    minimum height=2.5em]

\title{Variational Quantum State Diagonalization}

\author{Ryan LaRose}
\affiliation{Theoretical Division, Los Alamos National Laboratory, Los Alamos, NM 87545, USA.}
\affiliation{Department of Computational Mathematics, Science, and Engineering \& Department of Physics and Astronomy, Michigan State University, East Lansing, MI 48823, USA.}

\author{Arkin Tikku}
\affiliation{Theoretical Division, Los Alamos National Laboratory, Los Alamos, NM 87545, USA.}
\affiliation{Department of Physics, Blackett Laboratory, Imperial College London, Prince Consort Road, London SW7 2AZ, United Kingdom}

\author{\'Etude O'Neel-Judy}
\affiliation{Theoretical Division, Los Alamos National Laboratory, Los Alamos, NM 87545, USA.}

\author{Lukasz Cincio} 
\affiliation{Theoretical Division, Los Alamos National Laboratory, Los Alamos, NM 87545, USA.}

\author{Patrick J. Coles} 
\affiliation{Theoretical Division, Los Alamos National Laboratory, Los Alamos, NM 87545, USA.}

\begin{abstract}
Variational hybrid quantum-classical algorithms are promising candidates for near-term implementation on quantum computers. In these algorithms, a quantum computer evaluates the cost of a gate sequence (with speedup over classical cost evaluation), and a classical computer uses this information to adjust the parameters of the gate sequence. Here we present such an algorithm for quantum state diagonalization. State diagonalization has applications in condensed matter physics (e.g., entanglement spectroscopy) as well as in machine learning (e.g., principal component analysis). For a quantum state $\rho$ and gate sequence $U$, our cost function quantifies how far $ U\rho U^{\dagger}$ is from being diagonal. We introduce novel short-depth quantum circuits to quantify our cost. Minimizing this cost returns a gate sequence that approximately diagonalizes $\rho$. One can then read out approximations of the largest eigenvalues, and the associated eigenvectors, of $\rho$. As a proof-of-principle, we implement our algorithm on Rigetti's quantum computer to diagonalize one-qubit states and on a simulator to find the entanglement spectrum of the Heisenberg model ground state.
\end{abstract}

\maketitle


\section{Introduction}


The future applications of quantum computers, assuming that large-scale, fault-tolerant versions will eventually be realized, are manifold. From a mathematical perspective, applications include number theory \cite{shor1999polynomial}, linear algebra \cite{lloyd_quantum_2014-1, harrow2009quantum, rebentrost2018quantum}, differential equations \cite{leyton2008quantum, berry2014high}, and optimization \cite{farhi2014quantum}. From a physical perspective, applications include electronic structure determination \cite{peruzzo2014variational, kandala2017hardware} for molecules and materials and real-time simulation of quantum dynamical processes \cite{berry2015simulating} such as protein folding and photo-excitation events. Naturally, some of these applications are more long-term than others. Factoring and solving linear systems of equations are typically viewed as longer term applications due to their high resource requirements. On the other hand, approximate optimization and the determination of electronic structure may be nearer term applications, and could even serve as demonstrations of quantum supremacy in the near future \cite{preskill2012quantum, harrow2017quantum}.

A major aspect of quantum algorithms research is to make applications of interest more near term by reducing quantum resource requirements including qubit count, circuit depth, numbers of gates, and numbers of measurements. A powerful strategy for this purpose is algorithm hybridization, where a fully quantum algorithm is turned into a hybrid quantum-classical algorithm \cite{bravyi2016trading}. The benefit of hybridization is two-fold, both reducing the resources (hence allowing implementation on smaller hardware) as well as increasing accuracy (by outsourcing calculations to ``error-free'' classical computers).

Variational hybrid algorithms are a class of quantum-classical algorithms that involve minimizing a cost function that depends on the parameters of a quantum gate sequence. Cost evaluation occurs on the quantum computer, with speedup over classical cost evaluation, and the classical computer uses this cost information to adjust the parameters of the gate sequence. Variational hybrid algorithms have been proposed for Hamiltonian ground state and excited state preparation \cite{peruzzo2014variational, higgott2018variational, endo2018variational}, approximate optimization \cite{farhi2014quantum}, error correction \cite{johnson2017qvector}, quantum data compression \cite{romero2017quantum, Khoshaman_Vinci_Denis_Andriyash_Amin_2018}, quantum simulation \cite{li2017efficient, kokail2018self}, and quantum compiling \cite{Khatri_LaRose_Poremba_Cincio_Sornborger_Coles_2018}. A key feature of such algorithms is their near-term relevance, since only the subroutine of cost evaluation occurs on the quantum computer, while the optimization procedure is entirely classical, and hence standard classical optimization tools can be employed.

In this work, we consider the application of diagonalizing quantum states. In condensed matter physics, diagonalizing states is useful for identifying properties of topological quantum phases---a field known as entanglement spectroscopy \cite{li2008entanglement}. In data science and machine learning, diagonalizing the covariance matrix (which could be encoded in a quantum state \cite{giovannetti2008quantum, lloyd_quantum_2014-1}) is frequently employed for principal component analysis (PCA). PCA identifies features that capture the largest variance in one's data and hence allows for dimensionality reduction \cite{pearson1901liii}. 

Classical methods for diagonalization typically scale polynomially in the matrix dimension \cite{trefethen97}. Similarly, the number of measurements required for quantum state tomography---a general method for fully characterizing a quantum state---scales polynomially in the dimension. Interestingly, Lloyd et al.\ proposed a quantum algorithm for diagonalizing quantum states that can potentially perform exponentially faster than these methods \cite{lloyd_quantum_2014-1}. Namely, their algorithm, called quantum principal component analysis (qPCA), gives an exponential speedup for low-rank matrices. qPCA employs quantum phase estimation combined with density matrix exponentiation. These subroutines require a significant number of qubits and gates, making qPCA difficult to implement in the near term, despite its long-term promise.

Here, we propose a variational hybrid algorithm for quantum state diagonalization. For a given state $\rho$, our algorithm  is composed of three steps: (i) Train the parameters $\vec{\alpha}$ of a gate sequence $\U$ such that $\rhot = \Uopt \rho \Uopt \ad$ is approximately diagonal, where $\vec{\alpha}_{\text{opt}}$ is the optimal value of $\vec{\alpha}$ obtained (ii) Read out the largest eigenvalues of $\rho$ by measuring in the eigenbasis (i.e., by measuring $\rhot$ in the standard basis), and (iii) Prepare the eigenvectors associated with the largest eigenvalues. We call this the variational quantum state diagonalization (VQSD) algorithm. VQSD is a near-term algorithm with the same practical benefits as other variational hybrid algorithms. Employing a layered ansatz for $\U$ (where $p$ is the number of layers) allows one to obtain a hierarchy of approximations for the eigevalues and eigenvectors. We therefore think of VQSD as an approximate diagonalization algorithm.

We carefully choose our cost function $C$ to have the following properties: (i) $C$ is faithful (i.e, it vanishes if and only if $\rhot$ is diagonal), (ii) $C$ is efficiently computable on a quantum computer, (iii) $C$ has operational meanings such that it upper bounds the eigenvalue and eigenvector error (see Sec.~\ref{sec:vqsd}), and (iv) $C$ scales well for training purposes in the sense that its gradient does not vanish exponentially in the number of qubits. The precise definition of $C$ is given in Sec.~\ref{sec:vqsd} and involves a difference of purities for different states. To compute $C$, we introduce novel short-depth quantum circuits that likely have applications outside the context of VQSD.

To illustrate our method, we implement VQSD on Rigetti's 8-qubit quantum computer. We successfully diagonalize one-qubit pure states using this quantum computer. To highlight future applications (when larger quantum computers are made available), we implement VQSD on a simulator to perform entanglement spectroscopy on the ground state of the one-dimensional (1D) Heisenberg model composed of 12 spins.

Our paper is organized as follows. Section~\ref{sec:results} outlines the VQSD algorithm and presents its implementation. In Sec.~\ref{sec:discussion}, we give a comparison to the qPCA algorithm, and  we elaborate on future applications. Section \ref{sec:methods} presents our methods for quantifying diagonalization and for optimizing our cost function.

\section{Results} \label{sec:results}

\subsection{The VQSD Algorithm} \label{sec:vqsd}

\subsubsection{Overall Structure} \label{sec:vqsdstructure}

\begin{figure*}
    \centering
    \includegraphics[width = \linewidth]{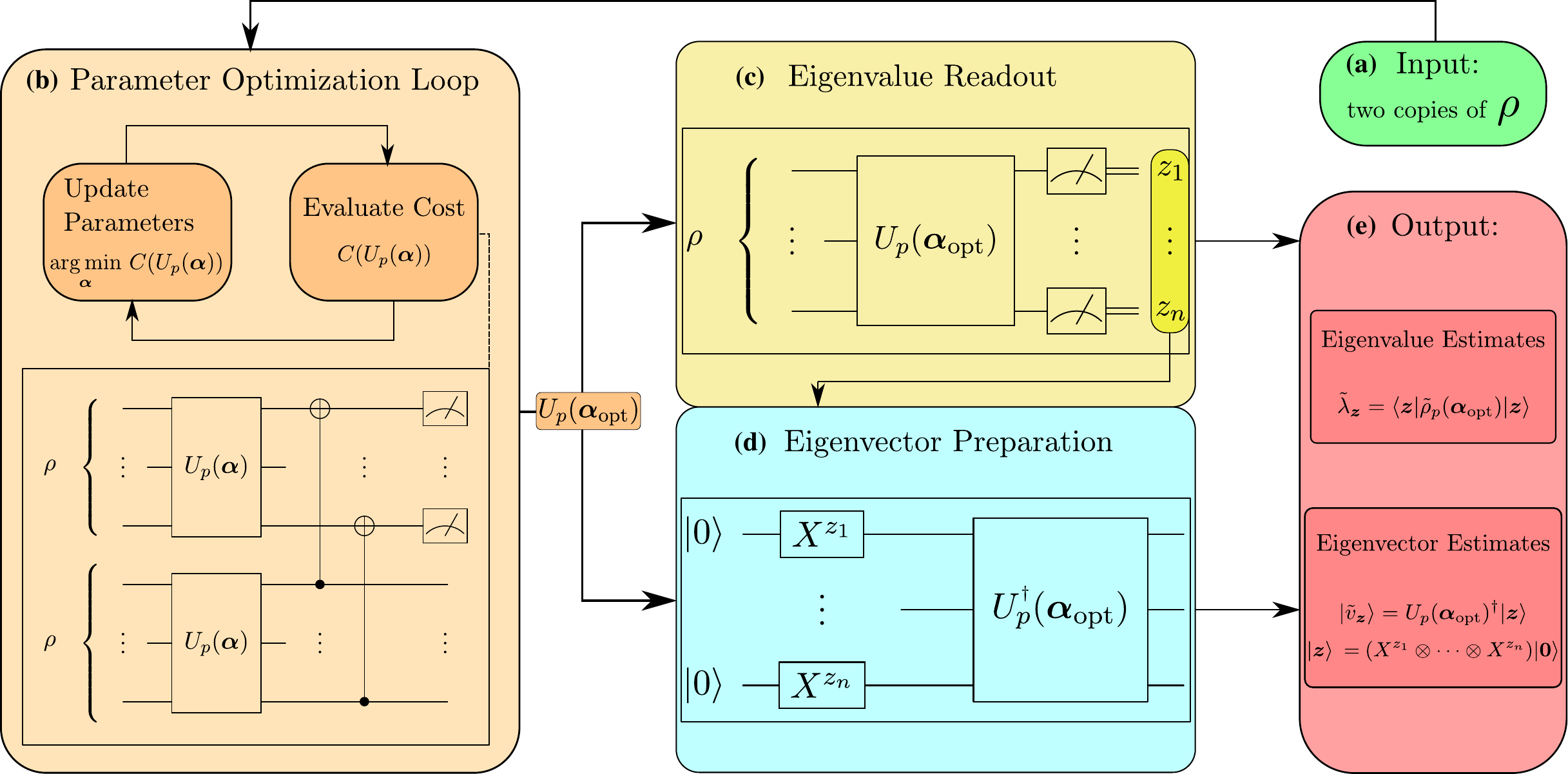}
    \caption{Schematic diagram showing the steps of the VQSD algorithm. (a) Two copies of quantum state $\rho$ are provided as an input.
    These states are sent to the parameter optimization loop (b) where a hybrid quantum-classical variational algorithm approximates the diagonalizing unitary $\Uopt$. Here, $p$ is a hyperparameter that dictates the quality of solution found. This optimal unitary is sent to the eigenvalue readout circuit (c) to obtain bitstrings $\vec{z}$, the frequencies of which provide estimates of the eigenvalues of $\rho$. Along with the optimal unitary $\Uopt$, these bitstrings are sent to the eigenvector preparation circuit (c) to prepare the eigenstates of $\rho$ on a quantum computer. Both the eigenvalues and eigenvectors are the outputs (d) of the VQSD algorithm. 
   }
    \label{fig:vqsd-overview}
\end{figure*}

Figure~\ref{fig:vqsd-overview} shows the structure of the VQSD algorithm. The goal of VQSD is to take, as its input, an $n$-qubit density matrix $\rho$ given as a quantum state and then output approximations of the $m$-largest eigenvalues and their associated eigenvectors. Here, $m$ will typically be much less than $2^n$, the matrix dimension of $\rho$, although the user is free to increase $m$ with increased algorithmic complexity (discussed below). The outputted eigenvalues will be in classical form, i.e., will be stored on a classical computer. In contrast, the outputted eigenvectors will be in quantum form, i.e., will be prepared on a quantum computer. This is necessary because the eigenvectors would have $2^n$ entries if they were stored on a classical computer, which is intractable for large $n$. Nevertheless, one can characterize important aspects of these eigenvectors with a polynomial number of measurements on the quantum computer.

Similar to classical eigensolvers, the VQSD algorithm is an approximate or iterative diagonalization algorithm. Classical eigenvalue algorithms are necessarily iterative, not exact \cite{footnote1}. Iterative algorithms are useful in that they allow for a trade-off between run-time and accuracy. Higher degrees of accuracy can be achieved at the cost of more iterations (equivalently, longer run-time), or short run-time can be achieved at the cost of lower accuracy. This flexibility is desirable in that it allows the user of the algorithm to dictate the quality of the solutions found.

The iterative feature of VQSD arises via a layered ansatz for the diagonalizing unitary. This idea similarly appears in other variational hybrid algorithms, such as the Quantum Approximate Optimization Algorithm \cite{farhi2014quantum}. Specifically, VQSD diagonalizes $\rho$ by variationally updating a parameterized unitary $\U$ such that
\begin{equation} \label{eqn:def-of-rho-prime}
    \rhot_p(\vec{\alpha}) := \U \rho \Udag
\end{equation}
is (approximately) diagonal at the optimal value $\vec{\alpha}_{\text{opt}}$. (For brevity we often write $\rhot$ for $\rhot_p(\vec{\alpha})$.) We assume a layered ansatz of the form
\begin{equation} \label{eqn:layered-ansatz-for-unitary}
    \U = L_1(\vec{\alpha}_1) L_2(\vec{\alpha}_2) \cdots L_p(\vec{\alpha}_p)\,.
\end{equation}
Here, $p$ is a hyperparameter that sets the number of layers $L_i(\vec{\alpha}_i)$, and each $\vec{\alpha}_i$ is a set of optimization parameters that corresponds to internal gate angles within the layer. The parameter $\vec{\alpha}$ in \eqref{eqn:def-of-rho-prime} refers to the collection of all $\vec{\alpha}_i$ for $i = 1, ..., p$. Once the optimization procedure is finished and returns the optimal parameters $\vec{\alpha}_{\text{opt}}$, one can then run a particular quantum circuit (shown in Fig.~\ref{fig:vqsd-overview}(c) and discussed below) $N_{\text{readout}}$ times to approximately determine the eigenvalues of $\rho$. The precision (i.e, the number of significant digits) of each eigenvalue increases with $N_{\text{readout}}$ and with the eigenvalue's magnitude. Hence for small $N_{\text{readout}}$ only the largest eigenvalues of $\rho$ will be precisely characterized, so there is a connection between $N_{\text{readout}}$ and how many eigenvalues, $m$, are determined. The hyperparameter $p$ is a refinement parameter, meaning that the accuracy of the eigensystem (eigenvalues and eigenvectors) typically increases as $p$ increases. We formalize this argument as follows.

Let $C$ denote our cost function, defined below in \eqref{eqn:cost-definition_overall}, which we are trying to minimize. In general, the cost $C$ will be non-increasing (i.e., will either decrease or stay constant) in $p$. One can ensure that this is true by taking the optimal parameters learned for $p$ layers as the starting point for the optimization of $p+1$ layers and by setting $\vec{\alpha}_{p+1}$ such that $L_{p+1}(\vec{\alpha}_{p+1})$ is an identity. This strategy also avoids barren plateaus \cite{mcclean2018barren, grant2019initialization} and helps to mitigate the problem of local minima, as we discuss in Appendix~\ref{app:lm} of Supplementary Material (SM).

Next, we argue that $C$ is closely connected to the accuracy of the eigensystem. Specifically, it gives an upper bound on the eigensystem error. Hence, one obtains an increasingly tighter upper bound on the eigensystem error as $C$ decreases (equivalently, as $p$ increases). To quantify eigenvalue error, we define
\begin{equation} \label{eqn:Delta}
 \Delta_{\lambda}:= \sum_{i=1}^{d} (\lambda_i - \tilde{\lambda}_i)^2\,,
\end{equation}
where $d=2^n$, and $\{ \lambda_i \}$ and $\{ \tilde{\lambda}_i \}$ are the true and inferred eigenvalues, respectively. Here, $i$ is an index that orders the eigenvalues in decreasing order, i.e., $\lambda_i \geq \lambda_{i+1}$ and $\tilde{\lambda}_i \geq \tilde{\lambda}_{i+1}$ for all $i\in \{1,...,d-1\}$. To quantify eigenvector error, we define
\begin{equation} \label{eqn:DeltaV}
 \Delta_{v}:= \sum_{i=1}^{d} \ip{\delta_i}{\delta_i}\,,\quad\text{with }\ket{\delta_i}=\rho \ket{\tilde{v}_i} - \tilde{\lambda}_i\ket{\tilde{v}_i}=\Pi_i^{\perp}\rho \ket{\tilde{v}_i}\,.
\end{equation}
Here, $\ket{\tilde{v}_i}$ is the inferred eigenvector associated with $\tilde{\lambda}_i$, and $\Pi_i^{\perp} = \id - \dya{\tilde{v}_i}$ is the projector onto the subspace orthogonal to $\ket{\tilde{v}_i}$. Hence, $\ket{\delta_i}$ is a vector whose norm quantifies the component of $\rho \ket{\tilde{v}_i}$ that is orthogonal to $\ket{\tilde{v}_i}$, or in other words, how far $\ket{\tilde{v}_i}$ is from being an eigenvector of $\rho$. 

As proven in Sec.~\ref{sec:circuits}, our cost function upper bounds the eigenvalue and eigenvector error up to a proportionality factor $\beta$,
\begin{equation} \label{eqn:UpperBoundOnDelta}
\Delta_{\lambda} \leq \beta C \,, \quad\text{and}\quad \Delta_{v} \leq \beta C \,.
\end{equation}
Because $C$ is non-increasing in $p$, the upper bound in \eqref{eqn:UpperBoundOnDelta} is non-increasing in $p$ and goes to zero if $C$ goes to zero.  

We remark that $\Delta_{v}$ can be interpreted as a weighted eigenvector error, where eigenvectors with larger eigenvalues are weighted more heavily in the sum. This is a useful feature since it implies that lowering the cost $C$ will force the eigenvectors with the largest eigenvalues to be highly accurate. In many applications, such eigenvectors are precisely the ones of interest. (See Sec.~\ref{sec:imp_Heis} for an illustration of this feature.)

The various steps in the VQSD algorithm are shown schematically in Fig.~\ref{fig:vqsd-overview}. There are essentially three main steps: (1) an optimization loop that minimizes the cost $C$ via back-and-forth communication between a classical and quantum computer, where the former adjusts $\vec{\alpha}$ and the latter computes $C$ for $U_p(\vec{\alpha})$, (2) a readout procedure for approximations of the $m$ largest eigenvalues, which involves running a quantum circuit and then classically analyzing the statistics, and (3)  a preparation procedure to prepare approximations of the eigenvectors associated with the $m$ largest eigenvalues. In the following subsections, we elaborate on each of these procedures.

\subsubsection{Parameter Optimization Loop} \label{subsec:param-optimization-loop}

\begin{figure}
    \centering
    \includegraphics[width=\columnwidth]{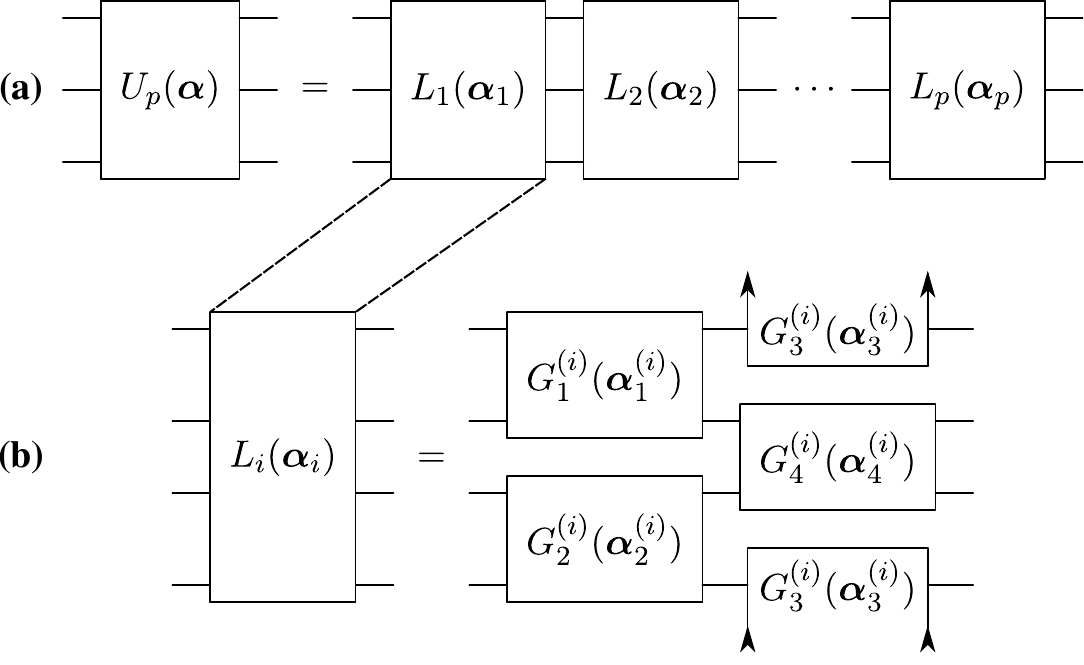}
    \caption{\textbf{(a)} Layered ansatz for the diagonalizing unitary $\U$. Each layer $L_i$, $i = 1, ..., p$, consists of a set of optimization parameters $\vec{\alpha}_i$. \textbf{(b)} The two-qubit gate ansatz for the $i$th layer, shown on four qubits. Here we impose periodic boundary conditions on the top/bottom edge of the circuit so that $G_3$ wraps around from top to bottom. Appendix~\ref{app:heisenberg} of SM discusses an alternative approach to the construction of $\U$, in which the ansatz is modified during the optimization process.
    }
    \label{fig:iterative-ansatz}
\end{figure}

Naturally, there are many ways to parameterize $U_p(\vec{\alpha})$. Ideally one would like the number of parameters to grow at most polynomially in both $n$ and $p$. Figure~\ref{fig:iterative-ansatz} presents an example ansatz that satisfies this condition. Each layer $L_i$ is broken down into layers of two-body gates that can be performed in parallel. These two-body gates can be further broken down into parameterized one-body gates, for example, with the construction in Ref.~\cite{VW04}. We discuss a different approach to parameterize $U_p(\vec{\alpha})$ in Appendix~\ref{app:heisenberg} of SM.

For a given ansatz, such as the one in Fig.~\ref{fig:iterative-ansatz}, parameter optimization involves evaluating the cost $C$ on a quantum computer for an initial choice of parameters and then modifying the parameters on a classical computer in an iterative feedback loop. The goal is to find
\begin{equation} \label{eqn:optimization-in-vqsd}
   \vec{\alpha}_{\text{opt}} := \underset{\vec{\alpha}}{\argmin} \ C(\U)\,.
\end{equation}
The classical optimization routine used for updating the parameters can involve either gradient-free or gradient-based methods. In Sec.~\ref{sec:optimization-methods}, we explore this further and discuss our optimization methods. 

In Eq.~\eqref{eqn:optimization-in-vqsd}, $C(\U)$ quantifies how far the state $\rhot_p(\vec{\alpha})$ is from being diagonal. There are many ways to define such a cost function, and in fact there is an entire field of research on \textit{coherence measures} that has introduced various such quantities \cite{baumgratz2014quantifying}. We aim for a cost that is efficiently computable with a quantum-classical system, and hence we consider a cost that can be expressed in terms of purities. (It is well known that a quantum computer can find the purity $\Tr(\sigma^2)$ of an $n$-qubit state $\sigma$ with complexity scaling only linearly in $n$, an exponential speedup over classical computation \cite{buhrman2001quantum, gottesman2001quantum}.) Two such cost functions, whose individual merits we discuss in Sec.~\ref{sec:circuits}, are
\begin{align}
\label{eqn:cost-definition_1}
  C_1(\U)  &= \Tr(\rho^2) - \Tr (\mathcal{Z}(\rhot)^2)\,,\\
\label{eqn:cost-definition_2}
  C_2(\U)  &= \Tr(\rho^2) - \frac{1}{n}\sum_{j=1}^n \Tr(\ZC_j(\rhot)^2)\,.
\end{align}
Here, $\ZC$ and $\ZC_j$ are quantum channels that dephase (i.e., destroy the off-diagonal elements) in the global standard basis and in the local standard basis on qubit $j$, respectively. Importantly, the two functions vanish under the same conditions:
\begin{align}
\label{eqn:costvanish}
  C_1(\U)  = 0 \iff C_2(\U)  = 0 \iff \rhot = \ZC(\rhot)\,.
\end{align}
So the global minima of $C_1$ and $C_2$ coincide and correspond precisely to unitaries $\U$ that diagonalize $\rho$ (i.e., unitaries such that $\rhot$ is diagonal).

As elaborated in Sec.~\ref{sec:circuits}, $C_1$ has operational meanings: it bounds our eigenvalue error, $C_1\geq \Delta_{\lambda}$, and it is equivalent to our eigenvector error, $C_1 = \Delta_{v}$. However, its landscape tends to be insensitive to changes in $\U$ for large $n$. In contrast, we are not aware of a direct operational meaning for $C_2$, aside from its bound on $C_1$ given by $C_2 \geq (1/n) C_1$. However, the landscape for $C_2$ is more sensitive to changes in $\U$, making it useful for training $\U$ when $n$ is large. Due to these contrasting merits of $C_1$ and $C_2$, we define our overall cost function $C$ as a weighted average of these two functions
\begin{align}
\label{eqn:cost-definition_overall}
 C(\U) = qC_1(\U) + (1-q) C_2(\U)\,,
\end{align}
where $q \in [0,1]$ is a free parameter that allows one to tailor the VQSD method to the scale of one's problem. For small $n$, one can set $q \approx 1$ since the landscape for $C_1$ is not too flat for small $n$, and, as noted above, $C_1$ is an operationally relevant quantity. For large $n$, one can set $q$ to be small since the landscape for $C_2$ will provide the gradient needed to train $\U$. The overall cost maintains the operational meaning in \eqref{eqn:UpperBoundOnDelta} with
\begin{equation}
\label{eqnBeta}
    \beta = n / (1 + q(n-1) )\,.
\end{equation}
Appendix \ref{app:qles1} illustrates the advantages of training with different values of $q$.

Computing $C$ amounts to evaluating the purities of various quantum states on a quantum computer and then doing some simple classical post-processing that scales linearly in $n$. This can be seen from Eqns.~\eqref{eqn:cost-definition_1} and \eqref{eqn:cost-definition_2}. The first term, $\Tr(\rho^2)$, in $C_1$ and $C_2$ is independent of $U_p(\vec{\alpha})$. Hence, $\Tr(\rho^2)$ can be evaluated outside of the optimization loop in Fig.~\ref{fig:vqsd-overview} using the Destructive Swap Test (see Sec.~\ref{sec:circuits} for the circuit diagram). Inside the loop, we only need to compute $\Tr(\ZC(\rhot)^2)$ and $\Tr(\ZC_j(\rhot)^2)$ for all $j$. Each of these terms are computed by first preparing two copies of $\rhot$ and then implementing quantum circuits whose depths are constant in $n$. For example, the circuit for computing $\Tr(\ZC(\rhot)^2)$ is shown in Fig.~\ref{fig:vqsd-overview}(b), and surprisingly it has a depth of only one gate. We call it the Diagonalized Inner Product (DIP) Test. The circuit for computing $\Tr(\ZC_j(\rhot)^2)$ is similar, and we call it the Partially Diagonalized Inner Product (PDIP) Test. We elaborate on both of these circuits in Sec.~\ref{sec:circuits}.

\subsubsection{Eigenvalue Readout} \label{subsec:spectrum-readout-and-eval-prep-circuits}

After finding the optimal diagonalizing unitary $\Uopt$, one can use it to readout approximations of the eigenvalues of $\rho$. Figure~\ref{fig:vqsd-overview}(c) shows the circuit for this readout. One prepares a single copy of $\rho$ and then acts with $\Uopt$ to prepare $\rhot_p(\vec{\alpha}_{\text{opt}})$. Measuring in the standard basis $\{\ket{\vec{z}}\}$, where $\vec{z} = z_1z_2 ... z_n$ is a bitstring of length $n$, gives a set of probabilities $\{\tilde{\lambda}_{\vec{z}}\}$ with
\begin{equation} \label{eqn:evalues1}
   \tilde{\lambda}_{\vec{z}} = \bra{\vec{z}}\rhot_p(\vec{\alpha}_{\text{opt}}) \ket{\vec{z}}\,.
\end{equation}
We take the $\tilde{\lambda}_{\vec{z}}$ as the inferred eigenvalues of $\rho$. We emphasize that the $\tilde{\lambda}_{\vec{z}}$ are the diagonal elements, not the eigenvalues, of $\rhot_p(\vec{\alpha}_{\text{opt}})$.

Each run of the circuit in Fig.~\ref{fig:vqsd-overview}(c) generates a bitstring $\vec{z}$ corresponding to the measurement outcomes. If one obtains $\vec{z}$ with frequency $f_{\vec{z}}$ for $N_{\text{readout}}$ total runs, then 
\begin{equation} \label{eqn:evalues2}
\tilde{\lambda}_{\vec{z}}^{\text{est}} = f_{\vec{z}}/N_{\text{readout}}
\end{equation}
gives an estimate for $\tilde{\lambda}_{\vec{z}}$. The statistical deviation of $\tilde{\lambda}_{\vec{z}}^{\text{est}}$ from $\tilde{\lambda}_{\vec{z}}$ goes with $1/\sqrt{N_{\text{readout}}}$. The relative error $\epsilon_{\vec{z}}$ (i.e., the ratio of the statistical error on $\tilde{\lambda}_{\vec{z}}^{\text{est}}$ to the value of $\tilde{\lambda}_{\vec{z}}^{\text{est}}$) then goes as
\begin{equation} \label{eqn:evalues3}
\epsilon_{\vec{z}} = \frac{1}{\sqrt{N_{\text{readout}}}\tilde{\lambda}_{\vec{z}}^{\text{est}}} =\frac{\sqrt{N_{\text{readout}}}}{f_{\vec{z}}}\,.
\end{equation}
This implies that events $\vec{z}$ with higher frequency $f_{\vec{z}}$ have lower relative error. In other words, the larger the inferred eigenvalue $\tilde{\lambda}_{\vec{z}}$, the lower the relative error, and hence the more precisely it is determined from the experiment. When running VQSD, one can pre-decide on the desired values of $N_{\text{readout}}$ and a threshold for the relative error, denoted $\epsilon_{\max}$. This error threshold $\epsilon_{\max}$ will then determine $m$, i.e., how many of the largest eigenvalues that get precisely characterized. So $m = m(N_{\text{readout}},\epsilon_{\max}, \{\tilde{\lambda}_{\vec{z}}\} )$ is a function of $N_{\text{readout}}$, $\epsilon_{\max}$, and the set of inferred eigenvalues $\{\tilde{\lambda}_{\vec{z}}\}$. Precisely, we take $m = |\vec{\tilde{\lambda}}^{\text{est}}|$ as the cardinality of the following set: 
\begin{equation} \label{eqn:evalues4}
\vec{\tilde{\lambda}}^{\text{est}} = \{\tilde{\lambda}_{\vec{z}}^{\text{est}} : \epsilon_{\vec{z}}\leq \epsilon_{\max}\}\,,
\end{equation}
which is the set of inferred eigenvalues that were estimated with the desired precision.

\subsubsection{Eigenvector Preparation} \label{subsec:eigenvector-preparation}

The final step of VQSD is to prepare the eigenvectors associated with the $m$-largest eigenvalues, i.e., the eigenvalues in the set in Eq.~\eqref{eqn:evalues4}. Let $\vec{Z} = \{\vec{z} : \tilde{\lambda}_{\vec{z}}^{\text{est}} \in \vec{\tilde{\lambda}}^{\text{est}}\}$ be the set of bitstrings $\vec{z}$ associated with the eigenvalues in $\vec{\tilde{\lambda}}^{\text{est}}$. (Note that these bitstrings are obtained directly from the measurement outcomes of the circuit in Fig.~\ref{fig:vqsd-overview}(c), i.e., the outcomes become the bitstring $\vec{z}$.) For each $\vec{z}\in \vec{Z}$, one can prepare the following state, which we take as the inferred eigenvector associated with our estimate of the inferred eigenvalue $\tilde{\lambda}_{\vec{z}}^{\text{est}}$,
\begin{align}
\label{eqn:evecs1}
\ket{\tilde{v}_{\vec{z}}}& = \Uopt\ad \ket{\vec{z}} \\
\label{eqn:evecs2}
&= \Uopt\ad (X^{z_1}\otimes \cdots \otimes X^{z_n})\ket{\vec{0}}\,.
\end{align}
The circuit for preparing this state is shown in Fig.~\ref{fig:vqsd-overview}(d). As noted in~\eqref{eqn:evecs2}, one first prepares $\ket{\vec{z}}$ by acting with $X$ operators raised to the appropriate powers, and then one acts with $\Uopt\ad$ to rotate from the standard basis to the inferred eigenbasis.

Once they are prepared on the quantum computer, each inferred eigenvector $\ket{\tilde{v}_{\vec{z}}}$ can be characterized by measuring expectation values of interest. That is, important physical features such as energy or entanglement (e.g., entanglement witnesses) are associated with some Hermitian observable $M$, and one can evaluate the expectation value $\mte{\tilde{v}_{\vec{z}}}{M}$ to learn about these features.

\subsection{Implementations} \label{sec:imp}

Here we present our implementations of VQSD, first for a one-qubit state on a cloud quantum computer to show that it is amenable to currently available hardware. Then, to illustrate the scaling to larger, more interesting problems, we implement VQSD on a simulator for the 12-spin ground state of the Heisenberg model. See Appendices~\ref{sec:app:details-on-vqsd-implementations} and \ref{app:heisenberg} of SM for further details. The code used to generate some of the examples presented here and in SM can be accessed from \cite{code}.

\subsubsection{One-Qubit State}

\begin{figure}[t]
    \centering
    \includegraphics[width=\columnwidth]{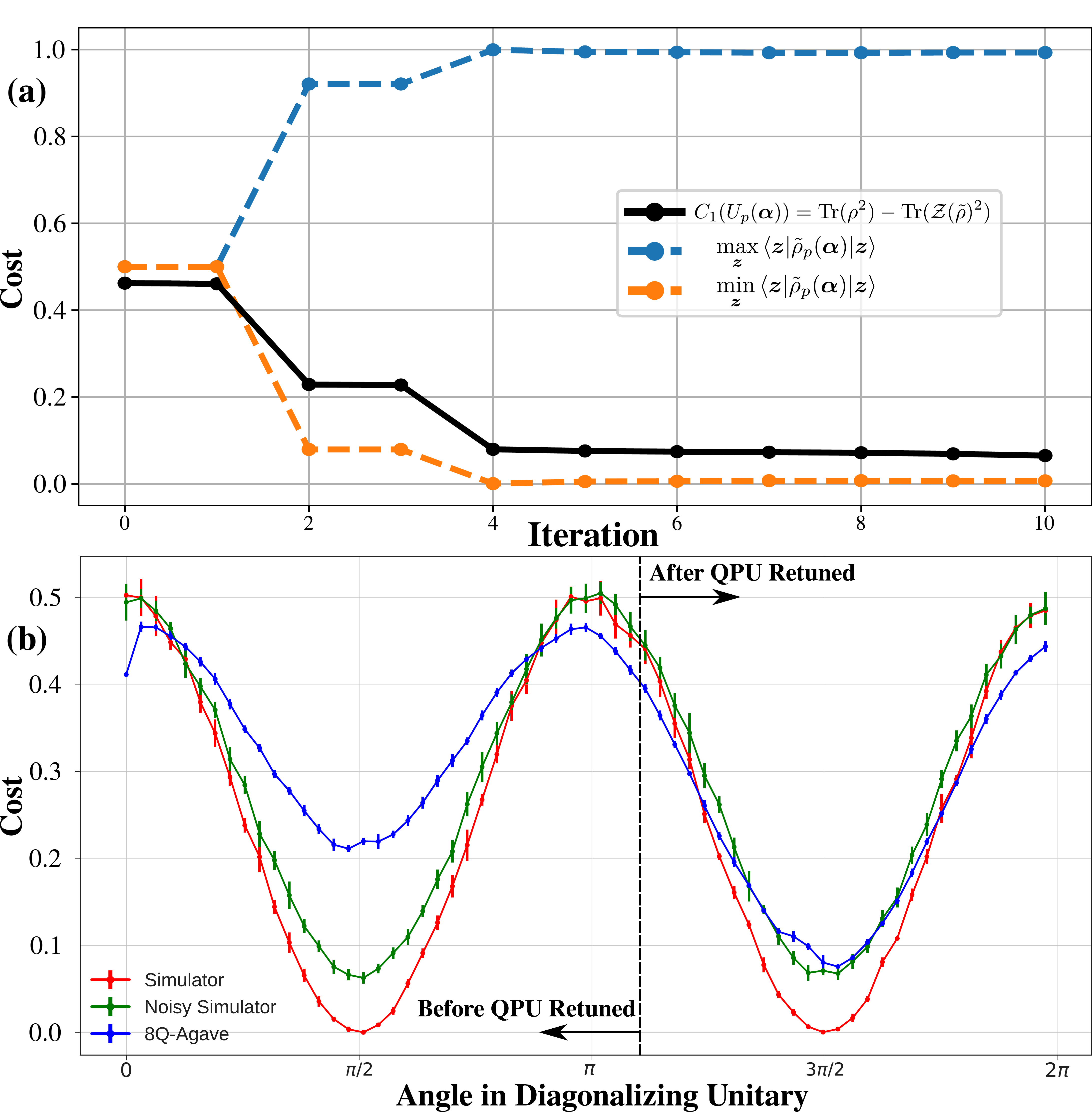}
    \caption{The VQSD algorithm run on Rigetti's 8Q-Agave quantum computer for $\rho = |+\>\<+|$.  (a) A representative run of the parameter optimization loop, using the Powell optimization algorithm (see Sec.~\ref{sec:optimization-methods} for details and Appendix~\ref{sec:app:details-on-vqsd-implementations} for data from additional runs). Cost versus iteration is shown by the black solid line. The dotted lines show the two inferred eigenvalues. After four iterations, the inferred eigenvalues approach $\{0,1\}$, as required for a pure state.  (b) The cost landscape on a noiseless simulator, Rigetti's noisy simulator, and Rigetti's quantum computer. Error bars show the standard deviation (due to finite sampling) of multiple runs. The local minima occur roughly at the theoretically predicted values of $\pi/2$ and $3\pi/2$. During data collection for this plot, the 8Q-Agave quantum computer retuned, after which its cost landscape closely matched that of the noisy simulator.}
    \label{fig:oneq-qpu-data}
\end{figure}

We now discuss the results of applying VQSD to the one-qubit plus state $\rho = |+\>\<+|$ on the 8Q-Agave quantum computer provided by Rigetti \cite{smith16quil}. Because the problem size is small ($n = 1$), we set $q = 1$ in the cost function \eqref{eqn:cost-definition_overall}. Since $\rho$ is a pure state, the cost function is
\begin{align}
C(\U) = C_1(\U)= 1 - \Tr (\mathcal{Z}(\rhot)^2) .
\end{align}
For $\U$, we take $p = 1$, for which the layered ansatz becomes an arbitrary single qubit rotation.

The results of VQSD for this state are shown in Fig.~\ref{fig:oneq-qpu-data}. In Fig.~\ref{fig:oneq-qpu-data}(a), the solid curve shows the cost versus the number of iterations in the parameter optimization loop, and the dashed curves show the inferred eigenvalues of $\rho$ at each iteration. Here we used the Powell optimization algorithm, see Section~\ref{sec:optimization-methods} for more details. As can be seen, the cost decreases to a small value near zero and the eigenvalue estimates simultaneously converge to the correct values of zero and one. Hence, VQSD successfully diagonalized this state. 

Figure~\ref{fig:oneq-qpu-data}(b) shows the landscape of the optimization problem on Rigetti's 8Q-Agave quantum computer, Rigetti's noisy simulator, and a noiseless simulator. Here, we varied the angle $\alpha$ in the diagonalizing unitary $U(\alpha) = R_x(\pi / 2) R_z(\alpha)$ and computed the cost at each value of this angle. The landscape on the quantum computer has local minima near the optimal angles $\alpha = \pi / 2, 3 \pi / 2$ but the cost is not zero. This explains why we obtain the correct eigenvalues even though the cost is nonzero in Fig.~\ref{fig:oneq-qpu-data}(a). The nonzero cost can be due to a combination of decoherence, gate infidelity, and measurement error. As shown in Fig.~\ref{fig:oneq-qpu-data}(b), the 8Q-Agave quantum computer retuned during our data collection, and after this retuning, the landscape of the quantum computer matched that of the noisy simulator significantly better.

\subsubsection{Heisenberg Model Ground State}\label{sec:imp_Heis}

\begin{figure}[t]
    \centering
    \includegraphics[width = \columnwidth]{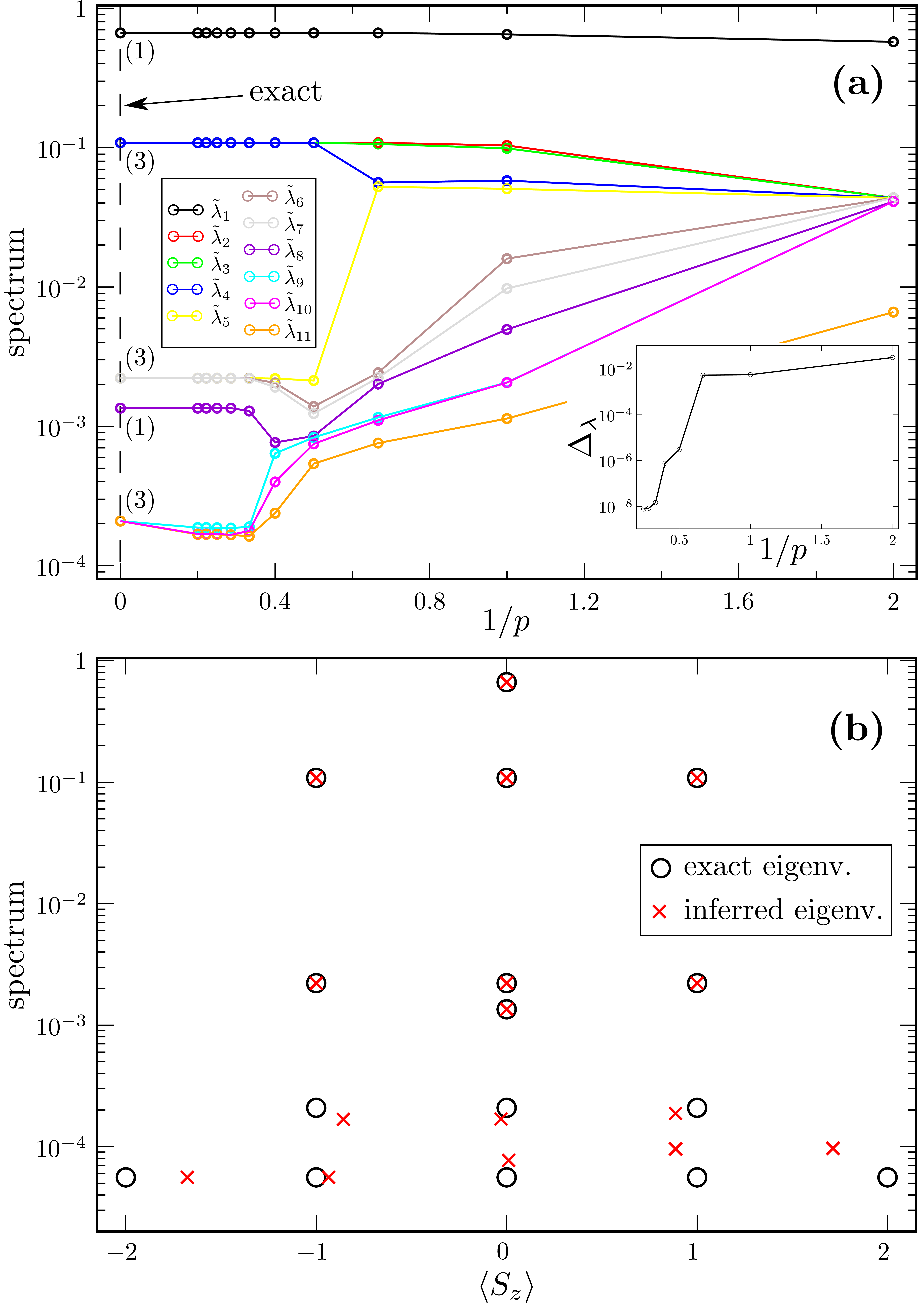}
    \caption{Implementing VQSD with a simulator for the ground state of the 1D Heisenberg model, diagonalizing a 4-spin subsystem of a chain of 8 spins. We chose $q=1$ for the cost in \eqref{eqn:cost-definition_overall} and employed a gradient-based method to find $\vec{\alpha}_{\text{opt}}$. (a) Largest inferred eigenvalues $\tilde{\lambda}_j$ versus $1/p$, where $p$ is the number of layers in our ansatz, which in this example takes half-integer values corresponding to fractions of layers shown in Fig.~\ref{fig:iterative-ansatz}. The exact eigenvalues are shown on the $y$-axis (along $1/p = 0$ line) with their degeneracy indicated in parentheses. One can see the largest eigenvalues converge to their correct values, including the correct degeneracies. Inset: overall eigenvalue error $\Delta_{\lambda}$ versus $1/p$. (b) Largest inferred eigenvalues resolved by the inferred $\avg{S_z}$ quantum number of their associated eigenvector, for $p=5$. The inferred data points (red X's) roughly agree with the theoretical values (black circles), particularly for the largest eigenvalues. Appendix~\ref{app:heisenberg} of SM discusses Heisenberg chain of 12 spins. }
    \label{fig:IsingModel}
\end{figure}

While current noise levels of quantum hardware limit our implementations of VQSD to small problem sizes, we can explore larger problem sizes on a simulator. An important application of VQSD is to study the entanglement in condensed matter systems, and we highlight this application in the following example.

Let us consider the ground state of the 1D Heisenberg model, the Hamiltonian of which is
\begin{align}
\label{eqn:HeisenbergH}
H =  \sum_{j=1}^{2n} \vec{S}^{(j)}\cdot \vec{S}^{(j+1)} \,,
\end{align}
with $\vec{S}^{(j)} = (1/2)(\sigma_x^{(j)}\hat{x}+\sigma_y^{(j)}\hat{y}+\sigma_z^{(j)}\hat{z}) $ and periodic boundary conditions, $\vec{S}^{(2n+1)} = \vec{S}^{(1)}$. Performing entanglement spectroscopy on the ground state $\ket{\psi}_{AB}$ involves diagonalizing the reduced state $\rho = \Tr_{B}(\dya{\psi}_{AB})$. Here we consider a total of 8 spins ($2n = 8$). We take $A$ to be a subset of 4 nearest-neighbor spins, and $B$ is the complement of $A$. 

The results of applying VQSD to the 4-spin reduced state $\rho$ via a simulator are shown in Fig.~\ref{fig:IsingModel}. Panel (a) plots the inferred eigenvalues versus the number of layers $p$ in our ansatz (see Fig.~\ref{fig:iterative-ansatz}). One can see that the inferred eigenvalues converge to their theoretical values as $p$ increases. Panel (b) plots the inferred eigenvalues resolved by their associated quantum numbers ($z$-component of total spin). This plot illustrates the feature we noted previously that minimizing our cost will first result in minimizing the eigenvector error for those eigenvectors with the largest eigenvalues. Overall our VQSD implementation returned roughly the correct values for both the eigenvalues and their quantum numbers. Resolving not only the eigenvalues but also their quantum numbers is important for entanglement spectroscopy \cite{li2008entanglement}, and clearly VQSD can do this.  

In Appendix~\ref{app:heisenberg} of SM we discuss an alternative approach employing a variable ansatz for $U_p(\vec{\alpha})$, and we present results of applying this approach to a 6-qubit reduced state of the 12-qubit ground state of the Heisenberg model.

\section{Discussion} \label{sec:discussion}

We emphasize that VQSD is meant for states $\rho$ that have either low rank or possibly high rank but low entropy $H(\rho) = -\Tr(\rho\log \rho)$. This is because the eigenvalue readout step of VQSD would be exponentially complex for states with high entropy. In other words, for high entropy states, if one efficiently implemented the eigenvalue readout step (with $N_{\text{readout}}$ polynomial in $n$), then very few eigenvalues would get characterized with the desired precision. In Appendix~\ref{app:complexity} of SM we discuss the complexity of VQSD for particular example states. 

Examples of states for which VQSD is expected to be efficient include density matrices computed from ground states of 1D, local, gapped Hamiltonians. Also, thermal states of some 1D systems in a many-body localized phase at low enough temperature are expected to be diagonalizable by VQSD. These states have rapidly decaying spectra and are eigendecomposed into states obeying a 1D area law \cite{hastings2007area, bauer2013area,grover2014}. This means that every eigenstate can be prepared by a constant depth circuit in alternating ansatz form \cite{bauer2013area}, and hence VQSD will be able to diagonalize it.

\subsection{Comparison to Literature} \label{sec:comparison-to-literature}

Diagonalizing quantum states with classical methods would require exponentially large memory to store the density matrix, and the matrix operations needed for diagonalization would be exponentially costly. VQSD avoids both of these scaling issues.

Another quantum algorithm that extracts the eigenvalues and eigenvectors of a quantum state is qPCA \cite{lloyd_quantum_2014-1}. Similar to VQSD, qPCA has the potential for exponential speedup over classical diagonalization for particular classes of quantum states. Like VQSD, the speedup in qPCA is contingent on $\rho$ being a low-entropy state. 

We performed a simple implementation of qPCA to get a sense for how it compares to VQSD, see Appendix~\ref{sec:appqpca} in SM for details. In particular, just like we did for Fig.~\ref{fig:oneq-qpu-data}, we considered the one-qubit plus state $\rho = \dya{+}$. We implemented qPCA for this state on Rigetti's noisy simulator (whose noise is meant to mimic that of their 8Q-Agave quantum computer). The circuit that we implemented applied one controlled-exponential-swap gate (in order to approximately exponentiate $\rho$, as discussed in \cite{lloyd_quantum_2014-1}). We employed a machine-learning approach \cite{cincio2018learning} to compile the controlled-exponential-swap gate into a novel short-depth gate sequence (see Appendix~\ref{sec:appqpca} in SM). With this circuit we inferred the two eigenvalues of $\rho$ to be approximately 0.8 and 0.2. Hence, for this simple example, it appears that qPCA gave eigenvalues that were slightly off from the true values of 1 and 0, while VQSD was able to obtain the correct eigenvalues, as discussed in Fig.~\ref{fig:oneq-qpu-data}.

\subsection{Future Applications}\label{sec:app}

Finally we discuss various applications of VQSD.

As noted in Ref.~\cite{lloyd_quantum_2014-1}, one application of quantum state diagonalization is benchmarking of quantum noise processes, i.e., quantum process tomography. Here one prepares the Choi state by sending half of a maximally entangled state through the process of interest. One can apply VQSD to the resulting Choi state to learn about the noise process, which may be particular useful for benchmarking near-term quantum computers. 

A special case of VQSD is variational state preparation. That is, if one applies VQSD to a pure state $\rho = \dya{\psi}$, then one can learn the unitary $U(\vec{\alpha})$ that maps $\ket{\psi}$ to a standard basis state. Inverting this unitary allows one to map a standard basis state (and hence the state $\ket{0}^{\otimes n}$) to the state $\ket{\psi}$, which is known as state preparation. Hence, if one is given $\ket{\psi}$ in quantum form, then VQSD can potentially find a short-depth circuit that approximately prepares $\ket{\psi}$. Variational quantum compiling algorithms that were very recently proposed \cite{Khatri_LaRose_Poremba_Cincio_Sornborger_Coles_2018, jones2018quantum} may also be used for this same purpose, and hence it would be interesting to compare VQSD to these algorithms for this special case. Additionally, in this special case one could use VQSD and these other algorithms as an error mitigation tool, i.e., to find a short-depth state preparation that achieves higher accuracy than the original state preparation.



In machine learning, PCA is a subroutine in supervised and unsupervised learning algorithms and also has many direct applications. PCA inputs a data matrix $X$ and finds a new basis such that the variance is maximal along the new basis vectors.  One can show that this amounts to finding the eigenvectors of the covariance matrix $E[X X^T]$ with the largest eigenvalues, where $E$ denotes expectation value. Thus PCA involves diagonalizing a positive-semidefinite matrix, $E[X X^T]$. Hence VQSD can perform this task provided one has access to QRAM \cite{giovannetti2008quantum} to prepare the covariance matrix as a quantum state. PCA can reduce the dimension of $X$ as well as filter out noise in data. In addition, nonlinear (kernel) PCA can be used on data that is not linearly separable. Very recent work by Tang \cite{tang2018quantum} suggests that classical algorithms could be improved for PCA of low-rank matrices, and potentially obtain similar scaling as qPCA and VQSD. Hence future work is needed to compare these different approaches to PCA. 


Perhaps the most important near-term application of VQSD is to study condensed matter physics. In particular, we propose that one can apply the variational quantum eigensolver \cite{peruzzo2014variational} to prepare the ground state of a many-body system, and then one can follow this with the VQSD algorithm to characterize the entanglement in this state. Ultimately this approach could elucidate key properties of condensed matter phases. In particular, VQSD allows for entanglement spectroscopy, which has direct application to the identification of topological order \cite{li2008entanglement}. Extracting both the eigenvalues and eigenvectors is useful for entanglement spectroscopy \cite{li2008entanglement}, and we illustrated this capability of VQSD in Fig.~\ref{fig:IsingModel}. Finally, an interesting future research direction is to check how the discrepancies in preparation of multiple copies affect the performance of the diagonalization.

\section{Methods} \label{sec:methods}

\subsection{Diagonalization Test Circuits}\label{sec:circuits}

Here we elaborate on the cost functions $C_1$ and $C_2$ and present short-depth quantum circuits to compute them.

\subsubsection{\texorpdfstring{$C_1$}{C1} and the DIP Test}

\begin{figure*}
    \centering
    \includegraphics[width=\textwidth]{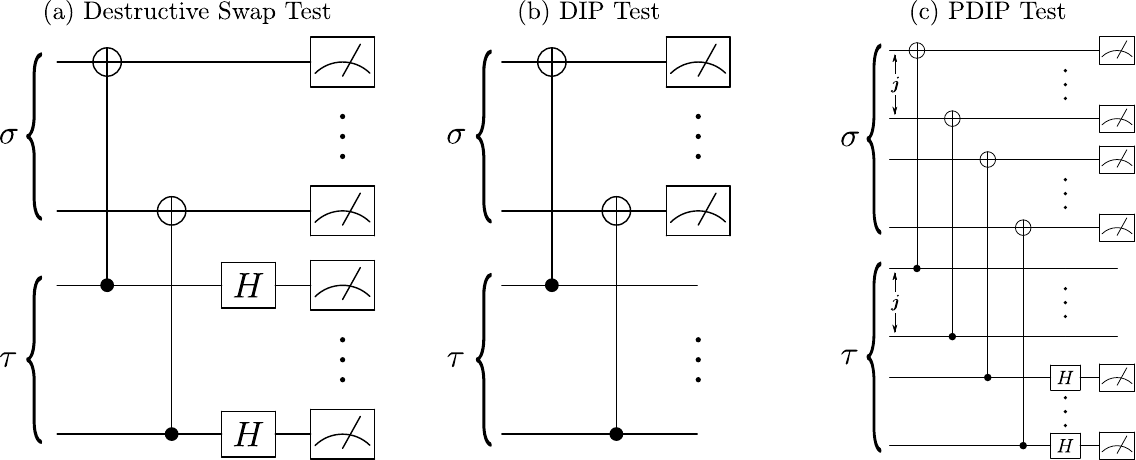}
    \caption{Diagonalization test circuits used in VQSD. (a) The Destructive Swap Test computes $\Tr(\sigma\tau)$ via a depth-two circuit. (b) The Diagonalized Inner Product (DIP) Test computes $\Tr(\ZC(\sigma)\ZC(\tau))$ via a depth-one circuit. (c) The Partially Diagonalized Inner Product (PDIP) Test computes $\Tr(\ZC_{\vec{j}}(\sigma)\ZC_{\vec{j}}(\tau))$ via a depth-two circuit, for a particular set of qubits $\vec{j}$. While the DIP test requires no postprocessing, the postprocessing for the Destructive Swap Test and the Partial DIP Test scales linearly in $n$.} 
    \label{fig:diptest}
\end{figure*}

The function $C_1$ defined in \eqref{eqn:cost-definition_1} has several intuitive interpretations. These interpretations make it clear that $C_1$ quantifies how far a state is from being diagonal. In particular, let $D_{\HS}(A,B):=\Tr\left((A-B)\ad (A-B)\right)$ denote the Hilbert-Schmidt distance. Then we can write 
\begin{align}
\label{eqn:cost-definition2}
  C_1 &= \min_{\sigma \in \DC}D_{\HS}(\rhot, \sigma)\\
\label{eqn:cost-definition2b}
  &= D_{\HS}(\tilde{\rho}, \mathcal{Z}(\rhot))\\
  \label{eqn:cost-definition3}
  &=  \sum_{\vec{z},\vec{z'}\neq \vec{z}} |\mted{\vec{z}}{\rhot}{\vec{z'}}|^2\,.
\end{align}
In other words, $C_1$ is (1) the minimum distance between $\rhot$ and the set of diagonal states $\DC$, (2) the distance from $\rhot$ to $\ZC(\rhot)$, and (3) the sum of the absolute squares of the off-diagonal elements of $\rhot$. 

$C_1$ can also be written as the eigenvector error in \eqref{eqn:DeltaV} as follows. For an inferred eigenvector $\ket{\tilde{v}_{\vec{z}}}$, we define $\ket{\delta_{\vec{z}}} = \rho\ket{\tilde{v}_{\vec{z}}} - \tilde{\lambda}_{\vec{z}}\ket{\tilde{v}_{\vec{z}}}$ and write the eigenvector error as 
\begin{align}
\ip{\delta_{\vec{z}}}{\delta_{\vec{z}}} &= \mte{\tilde{v}_{\vec{z}}}{\rho^2} + \tilde{\lambda}_{\vec{z}}^2 - 2 \tilde{\lambda}_{\vec{z}} \mte{\tilde{v}_{\vec{z}}}{\rho}\\ 
&=\mte{\tilde{v}_{\vec{z}}}{\rho^2} - \tilde{\lambda}_{\vec{z}}^2\,, 
\end{align}
since $\mte{\tilde{v}_{\vec{z}}}{\rho} = \tilde{\lambda}_{\vec{z}}$. Summing over all $\vec{z}$ gives
\begin{align}
\Delta_{v}= \sum_{\vec{z}}\ip{\delta_{\vec{z}}}{\delta_{\vec{z}}}&= \sum_{\vec{z}}\mte{\tilde{v}_{\vec{z}}}{\rho^2} - \tilde{\lambda}_{\vec{z}}^2\\
&= \Tr(\rho^2) - \Tr(\ZC(\rhot)^2) = C_1\,,
\label{eqnDeltavC1}
\end{align}
which proves the bound in (\ref{eqn:UpperBoundOnDelta}) for $q=1$.


In addition, $C_1$ bounds the eigenvalue error defined in \eqref{eqn:Delta}. Let $\vec{\tilde{\lambda}} = (\tilde{\lambda}_1, ..., \tilde{\lambda}_d)$ and $\vec{\lambda}= (\lambda_1, ..., \lambda_d)$ denote the inferred and actual eigenvalues of $\rho$, respectively, both arranged in decreasing order. In this notation we have
\begin{align}
\label{eqn:CostAndDelta1}
 \Delta_{\lambda} &= \vec{\lambda}\cdot \vec{\lambda}+ \vec{\tilde{\lambda}}\cdot \vec{\tilde{\lambda}} - 2\vec{\lambda}\cdot\vec{\tilde{\lambda}}\\
 \label{eqn:CostAndDelta2}
C_1&= \vec{\lambda}\cdot \vec{\lambda} -\vec{\tilde{\lambda}}\cdot \vec{\tilde{\lambda}} \\
\label{eqn:CostAndDelta3}
 &= \Delta_{\lambda} + 2(\vec{\lambda}\cdot\vec{\tilde{\lambda}} - \vec{\tilde{\lambda}}\cdot \vec{\tilde{\lambda}})\,.
\end{align}
Since the eigenvalues of a density matrix majorize its diagonal elements, $\vec{\lambda}\succ \vec{\tilde{\lambda}}$, and the dot product with an ordered vector is a Schur convex function, we have
\begin{align}
\label{eqn:CostAndDelta4}
\vec{\lambda}\cdot\vec{\tilde{\lambda}} \geq \vec{\tilde{\lambda}}\cdot \vec{\tilde{\lambda}}\,.
\end{align}
Hence from \eqref{eqn:CostAndDelta3} and \eqref{eqn:CostAndDelta4} we obtain the bound
\begin{equation} \label{eqn:CostAndDelta5}
\Delta_{\lambda} \leq C_1\,,
\end{equation}
which corresponds to the bound in \eqref{eqn:UpperBoundOnDelta} for the special case of $q=1$.

For computational purposes, we use the difference of purities interpretation of $C_1$ given in \eqref{eqn:cost-definition_1}. The $\Tr(\rho^2)$ term is independent of $\U$. Hence it only needs to be evaluated once, outside of the parameter optimization loop. It can be computed via the expectation value of the swap operator $S$ on two copies of $\rho$, using the identity 
\begin{align} \label{eqn:SwapTrick}
\Tr(\rho^2) = \Tr((\rho\otimes \rho) S)\,.
\end{align}
This expectation value is found with a depth-two quantum circuit that essentially corresponds to a Bell-basis measurement, with classical post-processing that scales linearly in the number of qubits \cite{garcia2013swap, cincio2018learning}. This is shown in Fig.~\ref{fig:diptest}(a). We call this procedure the Destructive Swap Test, since it is like the Swap Test, but the measurement occurs on the original systems instead of on an ancilla.

Similarly, the $\Tr (\mathcal{Z}(\rhot)^2)$ term could be evaluated by first dephasing $\rhot$ and then performing the Destructive Swap Test, which would involve a depth-three quantum circuit with linear classical post-processing. This approach was noted in Ref.~\cite{smith2017quantifying}. However, there exists a simpler circuit, which we call the Diagonalized Inner Product (DIP) Test. The DIP Test involves a depth-one quantum circuit with no classical post-processing. An abstract version of this circuit is shown in Fig.~\ref{fig:diptest}(b), for two states $\sigma$ and $\tau$. The proof that this circuit computes $\Tr(\ZC(\sigma) \ZC(\tau))$ is given in Appendix~\ref{app:cd} of SM. For our application we will set $\sigma = \tau = \rhot$, for which this circuit gives $\Tr(\ZC(\rhot)^2)$.

In summary, $C_1$ is efficiently computed by using the Destructive Swap Test for the $\Tr(\rho^2)$ term and the DIP Test for the  $\Tr(\ZC(\rhot)^2)$ term.

\subsubsection{\texorpdfstring{$C_2$}{C2} and the PDIP Test}

Like $C_1$, $C_2$ can also be rewritten in terms of of the Hilbert-Schmidt distance. Namely, $C_2$ is the average distance of $\rhot$ to each locally-dephased state $\ZC_j(\rhot)$:
\begin{align} \label{eqn:cost-definition_alt}
  C_2 &= \frac{1}{n}\sum_{j=1}^n D_{\HS}(\rhot, \ZC_j(\rhot))\,.
\end{align}
where $\ZC_j(\cdot) = \sum_{z} (\dya{z}_j\ot \id_{k\neq j})(\cdot)(\dya{z}_j\ot \id_{k\neq j})$. Naturally, one would expect that $C_2 \leq C_1$, since $\rhot$ should be closer to each locally dephased state than to the fully dephased state. Indeed this is true and can be seen from:
\begin{align} \label{eqn:cost-definition_alt2}
  C_2 = C_1 - \frac{1}{n}\sum_{j=1}^n \min_{\sigma \in \DC }D_{\HS}(\ZC_j(\rhot), \sigma)\,.
\end{align}
However, $C_1$ and $C_2$ vanish under precisely the same conditions, as noted in Eq.~\eqref{eqn:costvanish}. One can see this by noting that $C_2$ also upper bounds $(1/n)C_1$ and hence we have 
\begin{equation}
\label{eqnC2C1equivalence}
    C_2 \leq C_1 \leq n C_2\,.
\end{equation}
Combining the upper bound in \eqref{eqnC2C1equivalence} with the relations in \eqref{eqnDeltavC1} and \eqref{eqn:CostAndDelta5} gives the bounds in \eqref{eqn:UpperBoundOnDelta} with $\beta$ defined in \eqref{eqnBeta}. The upper bound in \eqref{eqnC2C1equivalence} is proved as follows. Let $\vec{z} = z_1 ... z_n$ and $\vec{z'} = z_1' ... z_n'$ be $n$-dimensional bitstrings. Let $\SC$ be the set of all pairs $(\vec{z},\vec{z'})$ such that $\vec{z} \neq \vec{z'}$, and let $\SC_j$ be the set of all pairs $(\vec{z},\vec{z'})$ such that  $z_j \neq z_j'$. Then we have $C_1 = \sum_{(\vec{z},\vec{z'})\in \SC} |\mted{\vec{z}}{\rhot}{\vec{z'}}|^2$, and
\begin{align}
    n C_2 &= \sum_{j=1}^n\sum_{(\vec{z},\vec{z'})\in \SC_j} |\mted{\vec{z}}{\rhot}{\vec{z'}}|^2 \\
    &\geq \sum_{(\vec{z},\vec{z'})\in \SC^U} |\mted{\vec{z}}{\rhot}{\vec{z'}}|^2 = C_1\,,
    \label{eqnC1C2offdiags}
\end{align}
where $\SC^U = \bigcup_{j=1}^n \SC_j $ is the union of all the $\SC_j$ sets. The inequality in \eqref{eqnC1C2offdiags} arises from the fact that the $\SC_j$ sets have non-trivial intersection with each other, and hence we throw some terms away when only considering the union $\SC^U$. The last equality follows from the fact that $\SC^U = \SC$, i.e, the set of all bitstring pairs that differ from each other ($\SC$) corresponds to the set of all bitstring pairs that differ for at least one element ($\SC^U$).

Writing $C_2$ in terms of purities, as in \eqref{eqn:cost-definition_2}, shows how it can be computed on a quantum computer. As in the case of $C_1$, the first term in \eqref{eqn:cost-definition_2} is computed with the Destructive Swap Test. For the second term in \eqref{eqn:cost-definition_2}, each purity $\Tr(\ZC_j(\rhot)^2)$ could also be evaluated with the Destructive Swap Test, by first locally dephasing the appropriate qubit. However, we present a slightly improved circuit to compute these purities that we call the Partially Diagonalized Inner Product (PDIP) Test. The PDIP Test is shown in Fig.~\ref{fig:diptest}(c) for the general case of feeding in two distinct states $\sigma$ and $\tau$ with the goal of computing the inner product between $\ZC_{\vec{j}}(\sigma)$ and $\ZC_{\vec{j}}(\tau)$. For generality we let $l$, with $0\leq l\leq n$, denote the number of qubits being locally dephased for this computation. If $l>0$, we define $\vec{j}= (j_1, \ldots , j_l)$ as a vector of indices that indicates which qubits are being locally dephased. The PDIP Test is a hybrid of the Destructive Swap Test and the DIP Test, corresponding to the former when $l = 0$ and the latter when $l=n$. Hence, it generalizes both the Destructive Swap Test and the DIP Test. Namely, the PDIP Test performs the DIP Test on the qubits appearing in $\vec{j}$ and performs the Destructive Swap Test on the qubits not appearing in $\vec{j}$. The proof that the PDIP Test computes $\Tr(\ZC_{\vec{j}}(\sigma)\ZC_{\vec{j}}(\tau))$, and hence $\Tr(\ZC_{\vec{j}}(\rhot)^2)$ when $\sigma = \tau = \rhot$, is given in Appendix~\ref{app:cd} of SM.

\subsubsection{\texorpdfstring{$C_1$}{C1} versus \texorpdfstring{$C_2$}{C2}}

Here we discuss the contrasting merits of the functions $C_1$ and $C_2$, hence motivating our cost definition in \eqref{eqn:cost-definition_overall}.

As noted previously, $C_2$ does not have an operational meaning like $C_1$. In addition, the circuit for computing $C_1$ is more efficient than that for $C_2$. The circuit in Fig.~\ref{fig:diptest}(b) for computing the second term in $C_1$ has a gate depth of one, with $n$ CNOT gates, $n$ measurements, and no classical post-processing. The circuit in Fig.~\ref{fig:diptest}(c) for computing the second term in $C_2$ has a gate depth of two, with $n$ CNOT gates, $n-1$ Hadamard gates, $2n-1$ measurements, and classical post-processing whose complexity scales linearly in $n$. So in every aspect, the circuit for computing $C_1$ is less complex than that for $C_2$. This implies that $C_1$ can be computed with greater accuracy than $C_2$ on a noisy quantum computer.

On the other hand, consider how the landscape for $C_1$ and $C_2$ scale with $n$. As a simple example, suppose $\rho = \dya{0} \otimes \cdots \otimes \dya{0}$. Suppose one takes a single parameter ansatz for $U$, such that $U(\theta) = R_X(\theta) \otimes \cdots \otimes R_X(\theta)$, where $R_X(\theta)$ is a rotation about the $X$-axis of the Bloch sphere by angle $\theta$. For this example,
\begin{align}\label{eqn:cost1example}
C_1(\theta) = 1 - \Tr(\ZC(\rhot)^2) = 1 - x(\theta)^n
\end{align}
where $x(\theta) = \Tr(\ZC(R_X(\theta)\dya{0}R_X(\theta)\ad)^2) = (1+\cos^2 \theta)/2$. If $\theta$ is not an integer multiple of $\pi$, then $x(\theta) < 1$, and $x(\theta)^n$ will be exponentially suppressed for large $n$. In other words, for large $n$, the landscape for $x(\theta)^n$ becomes similar to that of a delta function: it is zero for all $\theta$ except for multiples of $\pi$. Hence, for large $n$, it becomes difficult to train the unitary $U(\theta)$ because the gradient vanishes for most $\theta$. This is just an illustrative example, but this issue is general. Generally speaking, for large $n$, the function $C_1$ has a sharp gradient near its global minima, and the gradient vanishes when one is far away from these minima. Ultimately this limits $C_1$'s utility as a training function for large $n$.

In contrast, $C_2$ does not suffer from this issue. For the example in the previous paragraph,
\begin{align}\label{eqn:cost2example}
C_2(\theta) = 1 - x(\theta)\,,
\end{align}
which is independent of $n$. So for this example the gradient of $C_2$ does not vanish as $n$ increases, and hence $C_2$ can be used to train $\theta$. More generally, the landscape of $C_2$ is less barren than that of $C_1$ for large $n$. We can argue this, particularly, for states $\rho$ that have low rank or low entropy. The second term in \eqref{eqn:cost-definition_2}, which is the term that provides the variability with $\vec{\alpha}$, does not vanish even for large $n$, since (as shown in Appendix~\ref{sec:appProof} of SM):
\begin{align}\label{eqn:c2bound}
\Tr(\ZC_j(\rhot)^2) \geq 2^{-H(\rho) -1} \geq \frac{1}{2r}\,.
\end{align}
Here, $H(\rho) = -\Tr(\rho \log_2 \rho)$ is the von Neumann entropy, and $r$ is the rank of $\rho$. So as long as $\rho$ is low entropy or low rank, then the second term in $C_2$ will not vanish. Note that a similar bound does not exist for second term in $C_1$, which does tend to vanish for large $n$.


\subsection{Optimization Methods} \label{sec:optimization-methods}

Finding $\vec{\alpha}_{\text{opt}}$ in \eqref{eqn:optimization-in-vqsd} is a major component of VQSD. While many works have benchmarked classical optimization algorithms (e.g., Ref.~\cite{Rios2013}), the particular case of optimization for variational hybrid algorithms \cite{Guerreschi_Smelyanskiy_2017} is limited and needs further work \cite{mcclean2016theory}. Both gradient-based and gradient-free methods are possible, but gradient-based methods may not work as well with noisy data. Additionally, Ref.~\cite{mcclean2018barren} notes that gradients of a large class of circuit ansatze vanish when the number of parameters becomes large. These and other issues (e.g., sensitivity to initial conditions, number of function evaluations) should be considered when choosing an optimization method.

In our preliminary numerical analyses (see Appendix~\ref{sec:app:numerical-optimization} in SM), we found that the Powell optimization algorithm \cite{Powell_1978} performed the best on both quantum computer and simulator implementations of VQSD. This derivative-free algorithm uses a bi-directional search along each parameter using Brent's method. Our studies showed that Powell's method performed the best in terms of convergence, sensitivity to initial conditions, and number of correct solutions found. The implementation of Powell's algorithm used in this paper can be found in the open-source Python package SciPy Optimize \cite{scipy-opt}. Finally, Appendix~\ref{app:lm} of SM shows how our layered ansatz for $\U$ as well as proper initialization of $\U$ helps in mitigating the problem of local minima.

\section*{Data availability}
Data generated and analyzed during current study are available from the corresponding author upon reasonable request.

\section*{Code availability}
The code used to generate some of the examples presented here and in Supplementary Material can be accessed from \cite{code}.

\begin{acknowledgements}

We thank Rigetti for providing  access to their quantum computer. The views expressed in this paper are those of the authors and do not reflect those of Rigetti. RL, EO, and AT acknowledge support from the U.S. Department of Energy through a quantum computing program sponsored by the LANL Information Science \& Technology Institute. RL acknowledges support from an Engineering Distinguished Fellowship through Michigan State University. LC was supported by the U.S. Department of Energy through the J. Robert Oppenheimer fellowship. PJC was supported by the LANL ASC Beyond Moore's Law project. LC and PJC were also supported by the LDRD program at Los Alamos National Laboratory, by the U.S. Department of Energy (DOE), Office of Science, Office of Advanced Scientific Computing Research, and also by the U.S. DOE, Office of Science, Basic Energy Sciences, Materials Sciences and Engineering Division, Condensed Matter Theory Program.

\end{acknowledgements}

\section*{Author contribution}
RL, AT and LC implemented the algorithms and performed numerical analysis. LC and PJC designed the project. PJC proposed the cost function and proved the analytical results. RL, AT, \'EO-J, LC, and PJC contributed to data analysis, as well as writing and editing the final manuscript.

\section*{Competing interests}
The authors declare no competing interests.

\input{appendices.tex}

\end{document}

%% file: appendices.tex
\clearpage

\appendix
\setcounter{page}{1}
\renewcommand\thefigure{S.\arabic{figure}}
\setcounter{figure}{0}

\onecolumngrid
\begin{center}
\Large{ Supplementary Material for \\ ``Variational Quantum State Diagonalization''}
\end{center}

\twocolumngrid

\section{Details on VQSD Implementations} \label{sec:app:details-on-vqsd-implementations}

Here we provide further details on our implementations of VQSD in Sec.~\ref{sec:imp}. This includes further details about the optimization parameters as well as additional statistics for our runs on ns on the quantum computer.

\subsection{Optimization Parameters}

\begin{figure}[t!]
    \centering
    \includegraphics{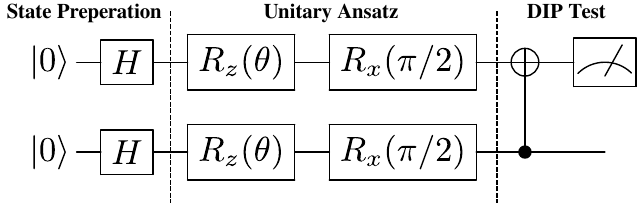}
    \caption{Circuit used to implement VQSD for $\rho=\dya{+}$ on Rigetti's 8Q-Agave quantum computer. Vertical dashed lines separate the circuit into logical components.}
    \label{fig:oneq-implementation-circuit}
\end{figure}

First, we discuss our implementation on a quantum computer (data shown in Fig.~\ref{fig:oneq-qpu-data}). Figure~\ref{fig:oneq-implementation-circuit} displays the circuit used for this implementation. This circuit is logically divided into three sections. First, we prepare two copies of the plus state $\rho = |+\>\<+| = H |0\> \<0| H$ by doing a Hadamard gate $H$ on each qubit. Next, we implement one layer of a unitary ansatz, namely $U(\theta) =  R_x(\pi / 2) R_z(\theta)$. This ansatz was chosen because each gate can be natively implemented on Rigetti's quantum computer. To simplify the search space, we restricted to one parameter instead of a universal one-qubit unitary. Last, we implement the DIP Test circuit, described in Fig.~\ref{fig:diptest}, which here consists of only one CNOT gate and one measurement.

For the parameter optimization loop, we used the Powell algorithm mentioned in Sec.~\ref{sec:optimization-methods}. This algorithm found the minimum cost in less than ten objective function evaluations on average. Each objective function evaluation (i.e., call to the quantum computer) sampled from 10,000 runs of the circuit in Fig.~\ref{fig:oneq-implementation-circuit}. As can be seen in Fig.~\ref{fig:oneq-qpu-data}(b), 10,000 runs was sufficient to accurately estimate the cost function \eqref{eqn:cost-definition_overall} with small variance. Because the problem size was small, we took $q = 1$ in \eqref{eqn:cost-definition_overall}, which provided adequate variability in the cost landscape.

Because of the noise levels in current quantum computers, we limited VQSD implementations on quantum hardware to only one-qubit states. Noise affects the computation in multiple areas. For example, in state preparation, qubit-specific errors can cause the two copies of $\rho$ to actually be different states. Subsequent gate errors (notably two-qubit gates), decoherence, and measurement errors prevent the cost from reaching zero even though the optimal value of $\theta$ is obtained. The effect of these various noise sources, and in particular the effect of discrepancies in preparation of two copies of $\rho$, will be important to study in future work.

Next, we discuss our VQSD implementation on a simulator (data shown in Fig.~\ref{fig:IsingModel}). For this implementation we again chose $q=1$ in our cost function. Because of the larger problem size (diagonalizing a 4-qubit state), we employed multiple layers in our ansatz, up to $p=5$. The simulator directly calculated the measurement probability distribution in the DIP Test, as opposed to determining the desired probability via sampling. This allowed us to use a gradient-based method to optimize our cost function, reducing the overall runtime of the optimization. Hence, our simulator implementation for the Heisenberg model demonstrated a future application of VQSD while alleviating the optimization bottleneck that is present for all variational quantum algorithms on large problem sizes, an area that needs further research \cite{mcclean2016theory}. We explore optimization methods further in Appendix~\ref{sec:app:numerical-optimization}.

\subsection{Additional statistics for the Quantum Computer Implementation}

\begin{figure}
    \centering
    \includegraphics[scale=0.30]{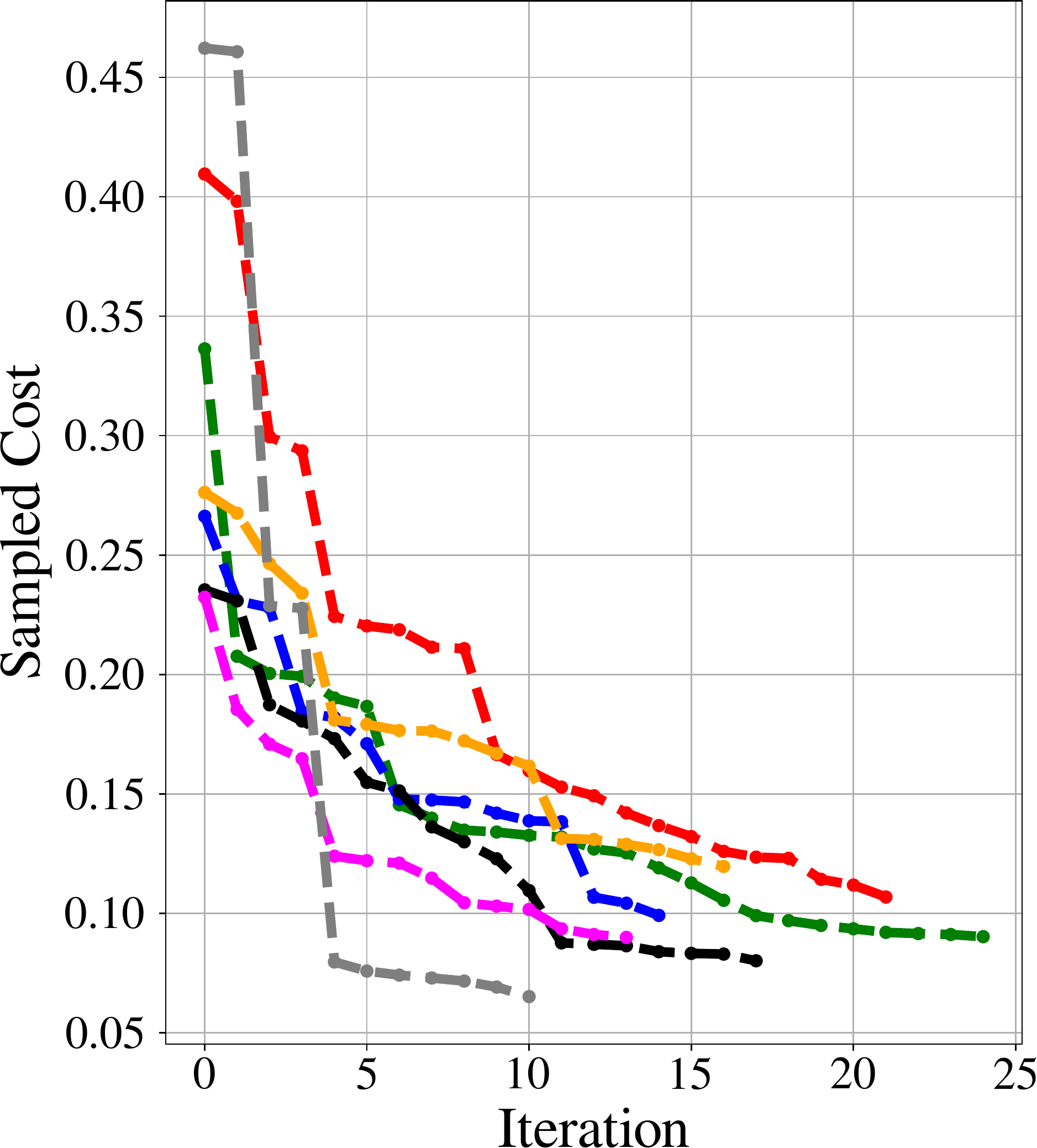}
    \caption{Cost vs iteration for all attempts of VQSD on Rigetti's 8Q-Agave computer for diagonalizing the plus state $\rho = |+\>\<+|$. Each of the seven curves represents a different independent run. Each run starts at a random initial angle and uses the Powell optimization algorithm to minimize the cost.}
    \label{fig:allcost}
\end{figure}

Here, we present statistics for several runs of the VQSD implementation run on Rigetti's 8Q-Agave quantum computer. One example plot of cost vs. iteration for diagonalizing the plus state $\rho = |+\>\<+|$ is shown in Figure~\ref{fig:oneq-qpu-data}(a). Here, we present all data collected for this implementation of VQSD, shown in Figure~\ref{fig:allcost}. The following table displays the final costs achieved as well the associated inferred eigenvalues.

\begin{table}[h!]
    \centering
    \begin{tabular}{|c|c|c|c|} \hline 
        VQSD Run    & $\min(C)$     & $\min(\tilde{\lambda_{\vec{z}}^{\text{est}}})$  & $\max(\tilde{\lambda_{\vec{z}}^{\text{est}}})$ \\ \hline 
        1 & 0.107 & 0.000        & 1.000       \\
        2 & 0.090 & 0.142        & 0.858        \\
        3 & 0.099 & 0.054        & 0.946       \\
        4 & 0.120 & 0.079        & 0.921       \\
        5 & 0.080 & 0.061        & 0.939       \\
        6 & 0.090 & 0.210        & 0.790       \\
        7 & 0.65 &  0.001        & 0.999       \\ \hline 
        \textbf{Avg.} & 0.093 & \textbf{0.078}       & \textbf{0.922}      \\
        \textbf{Std.} & 0.016 & 0.070         & 0.070      \\ \hline 
    \end{tabular}
    \caption{Minimum cost and eigenvalues achieved after performing the parameter optimization loop for seven independent runs of VQSD for the example discussed in Sec.~\ref{sec:imp}. The final two rows show average values and standard deviation across all runs.}
    \label{tab:my_label}
\end{table}

\section{Alternative Ansatz and the Heisenberg Model Ground State} \label{app:heisenberg}

In this Appendix, we describe a modification of the layered ansatz discussed in Section \ref{sec:vqsd}. Figure \ref{fig:iterative-ansatz} in the main text shows an example of a layered ansatz in which every layer has the same, fixed structure consisting of alternating two-qubit gates acting on nearest-neighbor qubits. The modified approach presented here may be useful in situations where there is no natural choice of the structure of the layered ansatz. 

Here, instead of working with a fixed structure for the diagonalizing unitary $U(\vec{\alpha})$, we allow it to vary during the optimization process. The algorithm used to update the structure of $U(\vec{\alpha})$ is probabilistic and resembles the one presented in \cite{cincio2018learning}.

In the examples studied here, the initial $U(\vec{\alpha})$ consists of a small number of random two-qubit gates with random supports (i.e. the qubits on which a gate acts). An optimization step involves minimizing the cost function by changing parameters $\vec{\alpha}$ as well as a small random change to the structure of $U(\vec{\alpha})$. This change to the structure typically amounts to a random modification of support for a limited number of gates. The new structure is accepted or rejected following the usual simulated annealing schemes. We refer the reader to Section II D of \cite{cincio2018learning} for further details on the optimization method.

The gate sequence representing $U(\vec{\alpha})$ is allowed to grow. If the algorithm described above cannot minimize the cost function for a specified number of iterations, an identity gate (spanned by new variational parameters) is randomly added to $U(\vec{\alpha})$. This step is similar in spirit to adding a layer to $U(\vec{\alpha})$ as discussed in Section \ref{sec:vqsd} of the main text.

\begin{figure}
    \centering
    \includegraphics[width=\columnwidth]{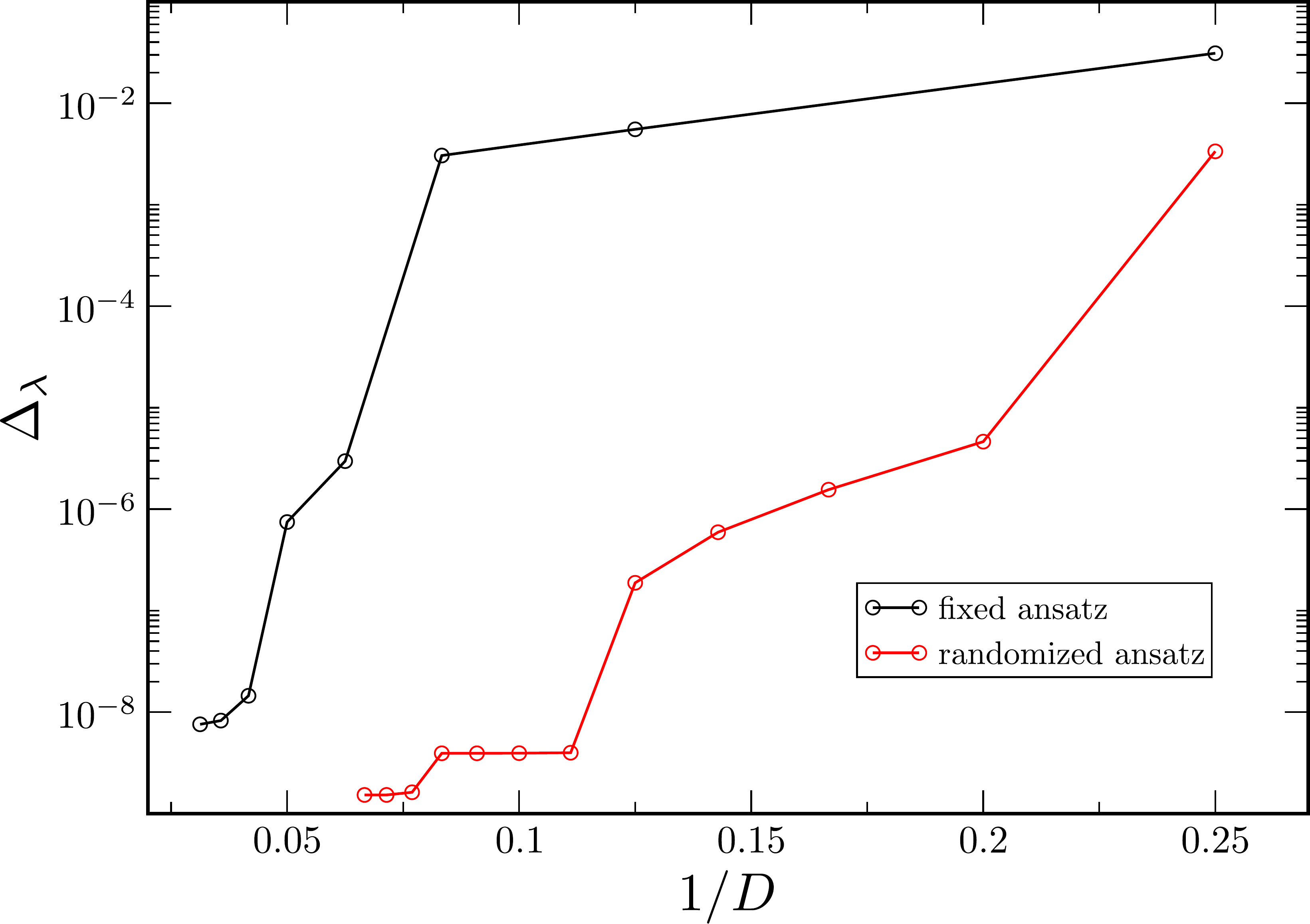}
    \caption{Comparison of two approaches to obtaining the diagonalizing unitary $U(\vec{\alpha})$: (i) based on a fixed layered ansatz shown in Fig. \ref{fig:iterative-ansatz} in the main text (black line) and (ii) based on random updates to the structure of $U(\vec{\alpha})$ (red line). The plot shows eigenvalue error $\Delta_\lambda$ versus $1/D$, where $D$ is the number of gates in $U(\vec{\alpha})$. For the same $D$, the second approach found a more optimal gate sequence. }
    \label{fig:fixed_vs_random}
\end{figure}

We compared the current method with the one based on the layered ansatz and found that it produced diagonalizing circuits involving significantly fewer gates. Figure \ref{fig:fixed_vs_random} shows the eigenvalue error $\Delta_\lambda$, defined in Eq.~(\ref{eqn:Delta}), as a function of $1/D$, where $D$ is the total number of gates of $U(\vec{\alpha})$. Here, VQSD is used to diagonalize a 4-qubit reduced state of the ground state of the one-dimensional Heisenberg model defined on 8 qubits, see Eq.~(\ref{eqn:HeisenbergH}). For every number of gates $D$, the current algorithm outperforms the one based on the fixed, layered ansatz. It finds a sequence of gates that results in a smaller eigenvalue error $\Delta_\lambda$. 

\begin{figure}
    \centering
    \includegraphics[width=\columnwidth]{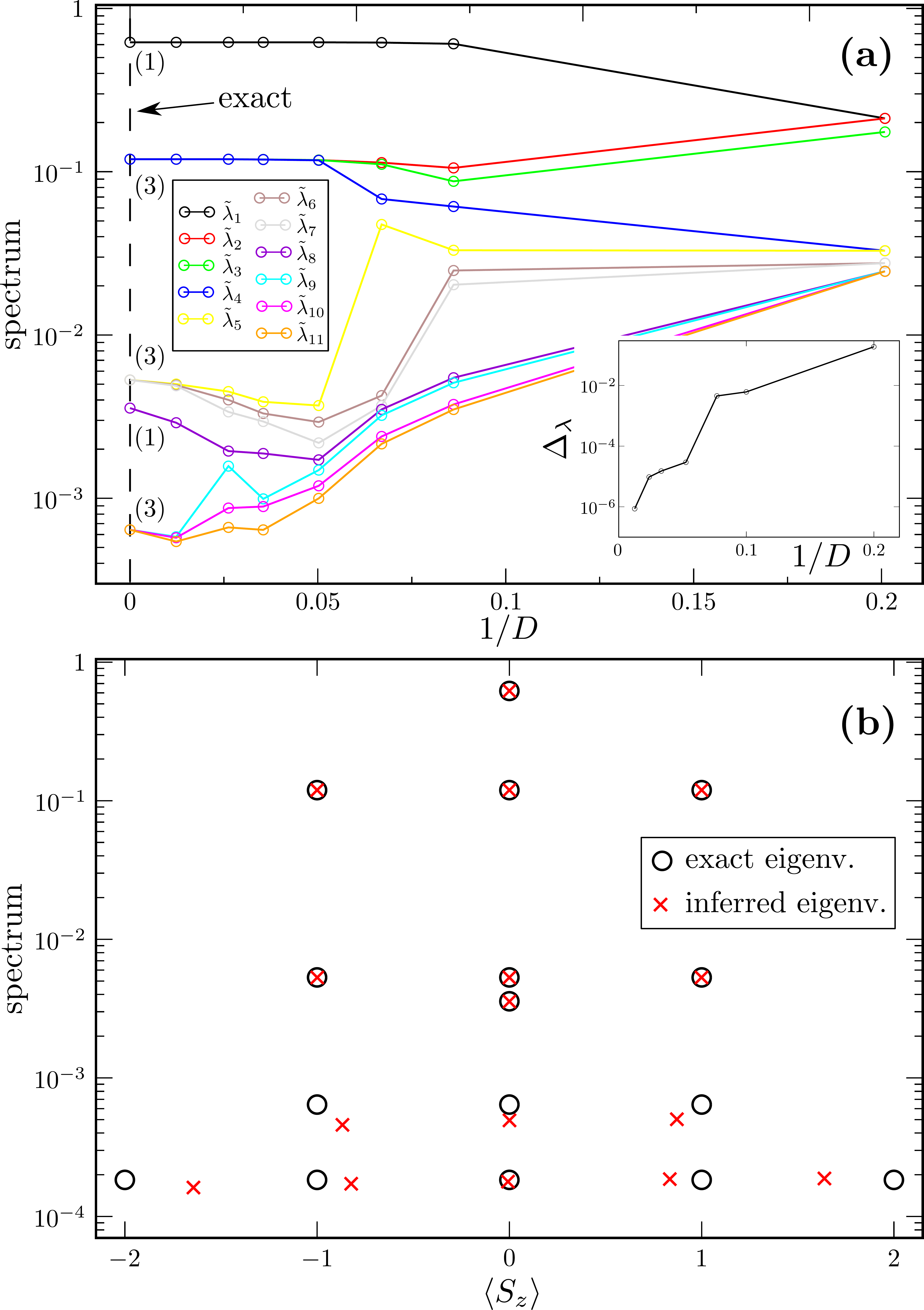}
    \caption{VQSD applied to the ground state of the Heisenberg model. Here we consider a 6-qubit reduced state $\rho$ of the 12-qubit ground state. {\bf (a)} Largest inferred eigenvalues $\tilde{\lambda}_j$ of $\rho$ as a function of $1/D$, where $D$ is the total number of gates in the diagonalizing unitary $U(\vec{\alpha})$. The inferred eigenvalues converge to their exact values shown along the $1/D = 0$ line recovering the correct degeneracy. Inset: Eigenvalue error $\Delta_\lambda$ as a function of $1/D$. {\bf (b)} The largest inferred eigenvalues $\tilde{\lambda}_j$ of $\rho$ resolved in the $\avg{S_z}$ quantum number. We find very good agreement between the inferred eigenvalues (red crosses) and the exact ones (black circles), especially for large eigenvalues. The data was obtained for $D=150$ gates.  }
    \label{fig:heis_Q6}
\end{figure}

Finally, we use the current optimization approach to find the spectrum of a 6-qubit reduced state $\rho$ of the 12-qubit ground state of a one-dimensional Heisenberg model. The results of performing VQSD on $\rho$ are shown in Fig. \ref{fig:heis_Q6}. Panel (a) shows the convergence of the 11 largest inferred eigenvalues $\tilde{\lambda}_j$ of $\rho$ to their exact values. We can see that the quality of the inferred eigenvalues increases quickly with the number of gates $D$ used in the diagonalizing unitary $U(\vec{\alpha})$. In panel (b), we show the dominant part of the spectrum of $\rho$ resolved in the $z$-component of the total spin. The results show that VQSD could be used to accurately obtain the dominant part of the spectrum of the density matrix together with the associated quantum numbers.

\section{Optimization and local minima} \label{app:lm}

In this Appendix we describe a strategy to avoid local minima that is used in the optimization algorithms throughout the paper and detailed in Appendix \ref{app:heisenberg}. We adapt the optimization involved in the diagonalization of the 6-qubit density matrix described in Appendix \ref{app:heisenberg} as an illustrative example. 

We note that the classical optimization problem associated with VQSD is potentially very difficult one. In the example studied in Appendix \ref{app:heisenberg} the diagonalizing unitary consisted of 150 two-qubit gates. This means that in order to find that unitary one has to optimize over at least $150 \cdot 13$ continuous parameters (every two-qubit gate is spanned by 15 parameters, but there is some reduction in the total number of parameters when two consecutive gates have overlapping supports). Initiated randomly, off-the-shelf techniques will most likely return suboptimal solution due to the presence of multiple local minima and the rough cost function landscape.

\begin{figure}[t!]
    \centering
    \includegraphics[width=\columnwidth]{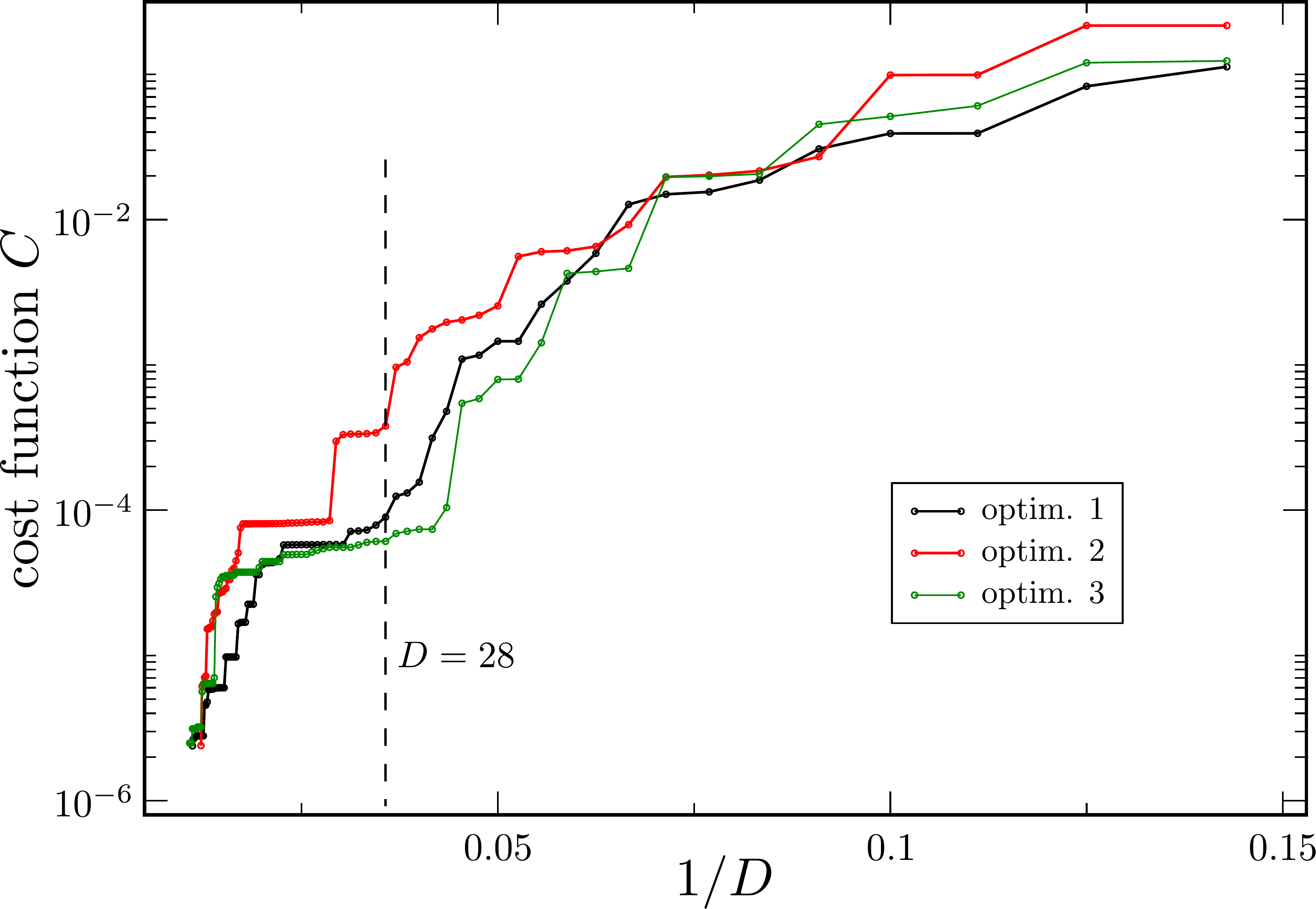}
    \caption{Cost function $C$ versus $1/D$ for three independent optimization runs. Here, $D$ is the total number of gates in the diagonalizing unitary $U_D(\vec{\alpha})$. Every optimization run got stuck at local minimum at some point during the minimization but thanks to the growth of the ansatz for $U_D(\vec{\alpha})$ described in the text, the predefined small value of $C$ was eventually attained. The data was obtained for a 6-qubit reduced state of the 12-qubit ground state of the Heisenberg model.}
    \label{fig:loc_min}
\end{figure}

Let $U_D(\vec{\alpha})$ denote a diagonalizing unitary that is built by $D$ two-qubit gates parametrized by $\vec{\alpha}$. Our optimization method begins with a shallow circuit consisting of few gates only. Since there is only a small number of variational parameters, the local minimum is quickly attained. After this initial step, the circuit that implements the unitary $U_D(\vec{\alpha})$ is grown by adding an identity gate (either randomly as discussed in this Appendix or by means of a layer of identity gates as presented in the main text). This additional gate contains new variational parameters that are initiated such that the unitary $U_D(\vec{\alpha}) = U_{D+1}(\vec{\alpha})$ and hence the value of the cost function are not changed. After the gate was added, the unitary $U_{D+1}(\vec{\alpha})$ contains more variational parameters which allows for further minimization of the cost function. In summary, the optimization of a deeper circuit $U_{D+1}(\vec{\alpha})$ is initialized by previously obtained $U_D(\vec{\alpha})$ as opposed to random initialization. What is more, even if the unitary $U_D(\vec{\alpha})$ was not the most optimal one for a given $D$, the growth of the circuit allows the algorithm to escape the local minimum and eventually find the global one, as illustrated by an example below and shown in Fig. \ref{fig:loc_min}. For a similar discussion, see \cite{grant2019initialization}.

To clarify the above analysis, let us consider an example of diagonalizing a 6-qubit reduced state of the 12-qubit ground state of the Heisenberg model, see Appendix \ref{app:heisenberg} for comparison. Figure \ref{fig:loc_min} shows the value of the cost function $C$ as a function of $1/D$ for three independent optimization runs. Each optimization was initialized randomly and we applied the same optimization scheme described above to each of them. We see that despite getting stuck in local minima, every optimization run managed to minimize the cost function to the predefined small value (which was set to $2 \cdot 10^{-6}$ in this example). For instance, at $D=28$, optimization run no. 2 clearly returns suboptimal solution (optimization run no. 3 gives lower cost function by a factor of 6) but after adding several identity gates, it manages to escape the local minimum and continue towards the global one.

\section{Optimization runs with various $q$ values}
\label{app:qles1}

\begin{figure*}
    \centering
    \includegraphics[width = \linewidth]{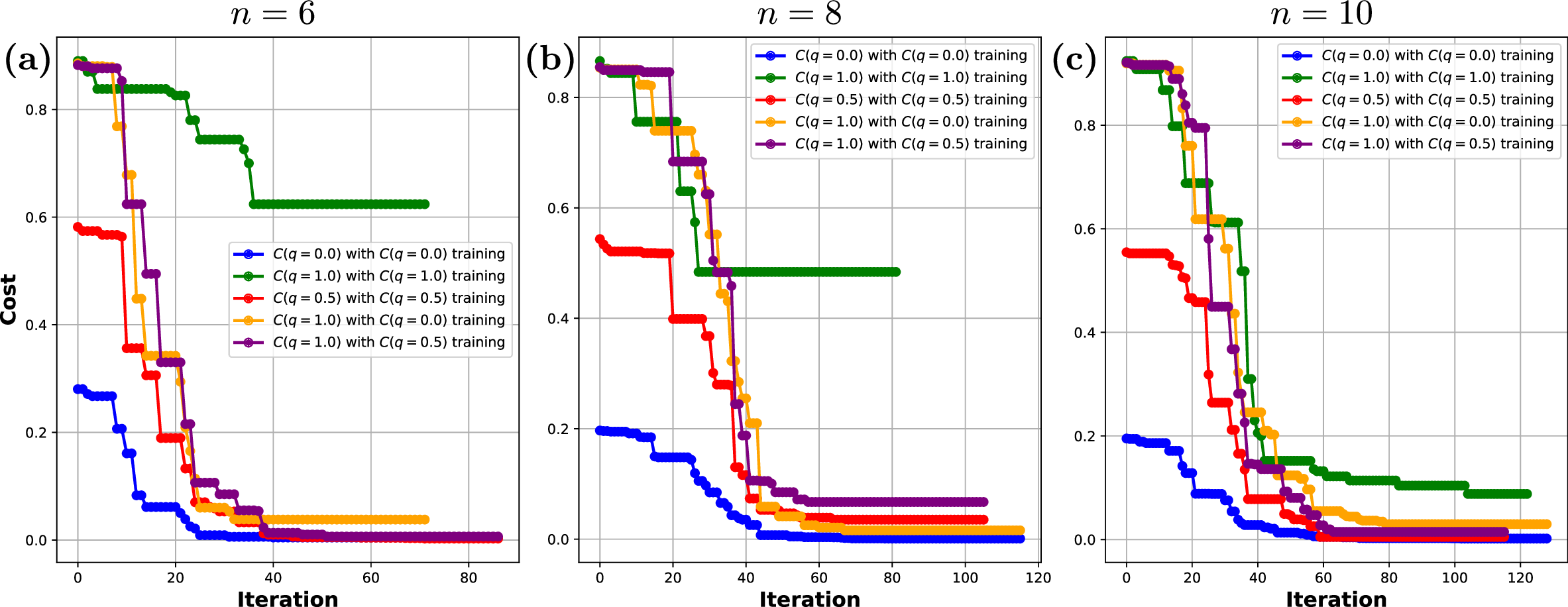}
    \caption{Cost versus iteration for different values of $q$, when $\rho$ is a tensor product of pure states on $n$ qubits. Here we consider (a) $n=6$, (b) $n=8$, and (a) $n=10$. We employed the COBYLA optimization method for training (see Appendix~\ref{sec:app:numerical-optimization} for discussion of this method). For each call to the quantum simulator (i.e., classical simulator of a quantum computer), we took 500 shots for statistics. The green, red, and blue curves respectively correspond to directly training the cost with $q=1$, $q=0.5$, and $q=0$. The purple and yellow curves respectively correspond to evaluating the $q=1$ cost for the angles $\vec{\alpha}$ obtained by training the $q=0.5$ and $q=0$ costs.
   }
    \label{fig:qlt1}
\end{figure*}

In this Appendix we present some numerical results for training our overall cost function for various values of $q$. Recall from Eq.~\eqref{eqn:cost-definition_overall} that $q$ is the weighting parameter that weights the contributions of $C_1$ and $C_2$ in the overall cost, as follows:
\begin{equation}
    C(\U) = q C_1(\U) + (1-q) C_2(\U)\,,
\end{equation}
where
\begin{align}
  C_1(\U)  &= \Tr(\rho^2) - \Tr (\mathcal{Z}(\rhot)^2)\,,\\
  C_2(\U)  &= \Tr(\rho^2) - \frac{1}{n}\sum_{j=1}^n \Tr(\ZC_j(\rhot)^2)\,.
\end{align}
As argued in Section~\ref{sec:results}, $C_1$ is operationally meaningful, while $C_2$ has a landscape that is more amendable to training when $n$ is large. In particular, one expects that for large $n$, the gradient of $C_1$ is sharp near the global minima but vanishes exponentially in $n$ away from these minima. In contrast, the gradient of $C_2$ is not expected to exponentially vanish as $n$ increases, even away from the minima.

Here, we numerically study the performance for different $q$ values for a simple example where $\rho$ is a tensor product of qubit pure states. Namely, we choose $\rho = \bigotimes_{j=1}^n V_j \dya{0} V_j\ad$, where $V_j = R_X(\theta_j)$ with $\theta_j$ randomly chosen. Such tensor product states are diagonalizable by a single layer ansatz: $U(\vec{\alpha}) = \bigotimes_{j=1}^n R_X(\alpha_j)$. We consider three different problem sizes: $n=6$, 8, and 10. Figure~\ref{fig:qlt1} shows our numerical results. 

Directly training the $C_1$ cost (corresponding to $q=1$) sometimes fails to find the global minimum. One can see this in Fig.~\ref{fig:qlt1}, where the green curve fails to fully reach zero cost. In contrast, the red and blue curves in Fig.~\ref{fig:qlt1}, which correspond to $q=0.5$ and $q=0$ respectively, approximately go to zero for large iterations. 

Even more interesting are the purple and yellow curves, which respectively correspond to evaluating the $C_1$ cost at the angles $\vec{\alpha}$ obtained from training the $q=0.5$ and $q=0$ costs. It is remarkable that both the purple and yellow curves perform better (i.e., achieve lower values) than the green curve. This implies that one can indirectly train the $C_1$ cost by training the $q=0.5$ or $q=0$ costs, and this indirect training performs better than directly training $C_1$. Since $C_1$ is operationally meaningful, this indirect training with $q<1$ is performing better in an operationally meaningful way.

We expect that direct training of $C_1$ will perform worse as $n$ increases, due to the exponential vanishing of the gradient of $C_1$. The particular runs shown in Fig.~\ref{fig:qlt1} do not show this trend, although this can be explained by the fact that the gradient of $C_1$ depends significantly on the initial values of the $\vec{\alpha}$, and indeed we saw large variability in the performance of the green curve even for a fixed $n$. Nevertheless, it is worth noting that we were always able to directly train $C_1$ (i.e., to make the green curve go to zero) for $n<6$, which is consistent with our expectations.

Overall, Fig.~\ref{fig:qlt1} provides numerical justification of the definition of our cost function as a weighted average, as in Eq.~(\ref{eqn:cost-definition_overall}). Namely, it shows that there is an advantage to choosing $q <1$.



\section{Comparison of Optimization Methods} \label{sec:app:numerical-optimization}

\begin{figure}
    \centering
    \includegraphics[width=\columnwidth]{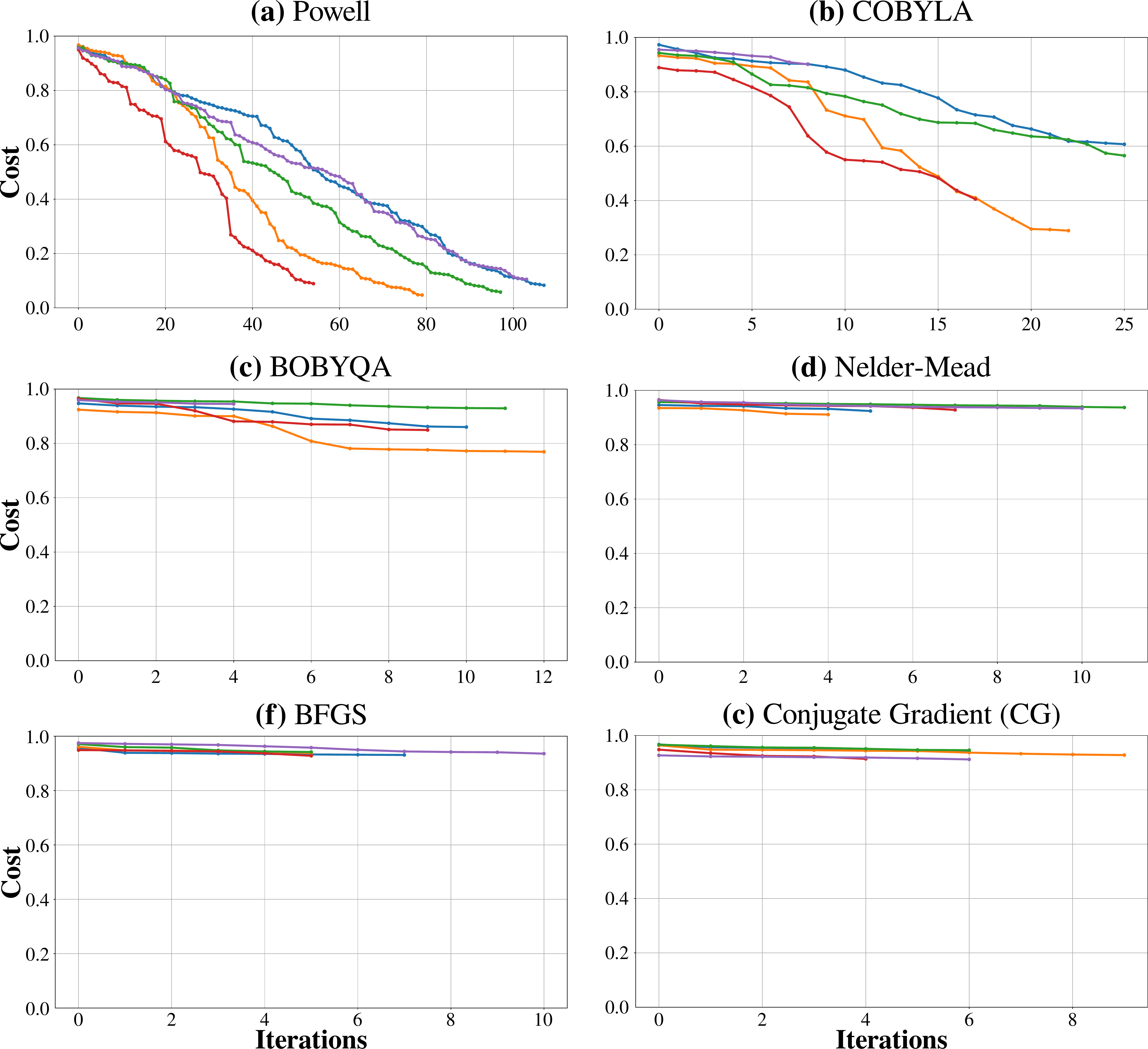}
    \caption{Optimization tests on six-qubit product states in the VQSD algorithm. Each plot shows a different optimization algorithm (described in main text) and curves on each plot show optimization attempts with different (random) initial conditions. Cost refers to the $C_1$ cost function ($q=1$ in \eqref{eqn:cost-definition_overall}), and each iteration is defined by a decrease in the cost function. As can be seen, the Powell algorithm is the most robust to initial conditions and provides the largest number of solved problem instances.}
    \label{fig:opt-tests}
\end{figure}

As emphasized previously, numerical optimization plays a key role in all variational hybrid algorithms, and further research in optimization methods is needed. In VQSD, the accuracy of the inferred eigenvalues are closely tied to the performance of the optimization algorithm used in the parameter optimization loop. This issue becomes increasingly important as one goes to large problem sizes (large $n$), where the number of parameters in the diagonalizing unitary becomes large.

Here, we compare the performance of six different optimization algorithms when used inside the parameter optimization loop of VQSD. These include Powell's algorithm \cite{Powell_1978}, Constrained Optimization BY Linear Approximation (COBYLA) \cite{powell_1998}, Bound Optimization BY Quadratic Approximation (BOBYQA) \cite{Powell2009TheBA}, Nelder-Mead \cite{Gao2012}, Broyden-Fletcher-Goldfarb-Shanno (BFGS) \cite{Nocedal_Wright_2006}, and conjugate gradient (CG) \cite{Nocedal_Wright_2006}. As mentioned in the main text, Powell's algorithm is a derivative-free optimizer that uses a bi-directional search along each parameter. The COBYLA and BOBYQA algorithms are both \textit{trust region} or \textit{restricted-step} methods, which approximate the objective function by a model function. The region where this model function is a good approximation is known as the trust region, and at each step the optimizer attempts to expand the trust region. The Nelder-Mead algorithm is a simplex method useful for derivative-free optimization with smooth objective functions. Lastly, the BFGS and CG algorithms are both gradient-based. The BFGS method is a quasi-Newton method that uses first derivatives only, and the CG method uses a nonlinear conjugate gradient descent. The implementations used in our study can be found in the open-source Python package SciPy Optimize \cite{scipy-opt} and in Ref.~\cite{Cartis_Fiala_Marteau_Roberts_2018}.

For this study, we take the input state $\rho$ to be a six-qubit pure product state:
\begin{equation}
\rho = \bigotimes_{j=1}^6\dya{\psi_j}\,,\quad\text{where}\quad \ket{\psi_j} =V_j\ket{0}\,. 
\end{equation}
Here, the state preparation unitary is 
\begin{equation}
    V_j = R_x(\alpha_x^{(j)}) R_y(\alpha_y^{(j)}) R_z(\alpha_z^{(j)})
\end{equation}
where the angles $(\alpha_x^{(j)}, \alpha_y^{(j)}, \alpha_z^{(j)})$ are randomly chosen. 

Using each algorithm, we attempt to minimize the cost by adjusting 36 parameters in one layer of the unitary ansatz in Fig.~\ref{fig:iterative-ansatz}. For fairness of comparison, only the objective function and initial starting point were input to each algorithm, i.e., no special options such as constraints, bounds, or other information was provided. The results of this study are shown in Fig.~\ref{fig:opt-tests} and Table~\ref{tab:relative-runtimes}.

Figure~\ref{fig:opt-tests} shows cost versus iteration for each of the six algorithms. Here, we define one iteration by a call to the objective function in which the cost decreases. In particular, the number of iterations is different than the number of cost function evaluations (see Table~\ref{tab:relative-runtimes}), which is not set a priori but rather determined by the optimizer. Plotting cost per each function evaluation would essentially produce a noisy curve since the optimizer is trying many values for the parameters. Instead, we only plot the cost for each parameter update in which the cost decreases. Both the number of iterations, function evaluations, and overall runtime are important features of the optimizer.

\begin{table}
    \centering
    \begin{tabular}{|c|c|c|c|c|c|c|c|} \hline 
        Alg.    &    Powell     & COBYLA    & BOBYQA    & Nelder-Mead       & BFGS      & CG        \\ \hline 
        r.r.    &   13.20       & 1         & 2.32      & 23.65             & 3.83      & 2.89      \\ \hline 
        f.ev.   & 4474          & 341       & 518       & 7212              & 1016      & 1045      \\ \hline 
    \end{tabular}
    \caption{Relative average run-times (r.r.) and absolute number of function evaluations (f.ev.) of each optimization algorithm (Alg.) used for the data obtained in Fig.~\ref{fig:opt-tests}. For example, BOBYQA took 2.32 times as long to run on average than COBYLA, which took the least time to run out of all algorithms. Absolute run-times depend on a variety of factors and computer performance. For reference, the COBYLA algorithm takes approximately one minute for this problem on a laptop computer. The number of cost function evaluations used (related to run-time but also dependent on the method used by the optimizer) is shown in the second row.}
    \label{tab:relative-runtimes}
\end{table}

In this study, as well as others, we found that the Powell optimization algorithm provides the best performance in terms of lowest minimum cost achieved, sensitivity to initial conditions, and fraction of correct solutions found. The trust-region algorithms COBYLA and BOBYQA were the next best methods. In particular, although the Powell algorithm consistently obtained lower minimum costs, the COBYLA method ran thirteen times faster on average (see Table~\ref{tab:relative-runtimes}). Indeed, both trust region methods provided the shortest runtime. The gradient-based methods BFGS and CG had comparable run-times but were unable to find any minima. Similarly, the Nelder-Mead simplex algorithm was unable to find any minima. This method also had the longest average run-time of all algorithms tested.

This preliminary analysis suggests that the Powell algorithm is the best method for VQSD. For other variational quantum algorithms, this may not necessarily be the case. In particular, we emphasize that the optimization landscape is determined by both the unitary ansatz and the cost function definition, which may vary drastically in different algorithms. While we found that gradient-based methods did not perform well for VQSD, they may work well for other applications. Additionally, optimizers that we have not considered here may also provide better performance. We leave these questions to further work.

\section{Complexity for particular examples}\label{app:complexity}

\subsection{General Complexity Remarks} \label{subsec:general-complexity-remarks}

In what follows we discuss some simple examples of states to which one might apply VQSD. There are several aspects of complexity to keep in mind when considering these examples, including:

(C1) The gate complexity of the unitary that diagonalizes $\rho$. (It is worth remarking that approximate diagonalization might be achieved with a less complex unitary than exact diagonalization.)

(C2) The complexity of searching through the search space to find the diagonalizing unitary.

(C3) The statistical complexity associated with reading out the eigenvalues. 

Naturally, (C1) is related to (C2). However, being efficient with respect to (C1) does not guarantee that (C2) is efficient.

\subsection{Example States} \label{sec:example-states}

In the simplest case, suppose $\rho = \dya{\psi_1}\otimes \cdots \otimes \dya{\psi_n}$ is a tensor product of pure states. This state can be diagonalized by a depth-one circuit $U = U_1 \otimes \cdots \otimes U_n$ composed of $n$ one-qubit gates (all done in parallel). Each $U_j$ diagonalizes the associated $\dya{\psi_j}$ state. Searching for this unitary within our ansatz can be done by setting $p=1$, i.e., with a single layer $L_1$ shown in Fig.~\ref{fig:iterative-ansatz}. A single layer is enough to find the unitary that exactly diagonalizes $\rho$ in this case. Hence, for this example, both complexities (C1) and (C2) are efficient. Finally, note that the eigenvalue readout, (C3), is efficient because there is only one non-zero eigenvalue. Hence, $\tilde{\lambda}_{\vec{z}}^{\text{est}}\approx 1$ and $\epsilon_{\vec{z}}\approx 1/\sqrt{N_{\text{readout}}}$ for this eigenvalue. This implies that $N_{\text{readout}}$ can be chosen to be constant, independent of $n$, in order to accurately characterize this eigenvalue.

A generalization of product states are classically correlated states, which have the form
\begin{equation}
    \rho = \sum_{\vec{z}} p_{\vec{z}} \dya{b^{(1)}_{z_1}} \otimes \cdots \otimes \dya{b^{(n)}_{z_n}}
\end{equation}
where $\{\ket{b^{(j)}_{0}},\ket{b^{(j)}_{1}} \}$ form an orthonormal basis for qubit~$j$. Like product states, classically correlated states can be diagonalized with a depth-one circuit composed of one-body unitaries. Hence (C1) and (C2) are efficient for such states. However, the complexity of eigenvalue readout depends on the $\{p_{\vec{z}}\}$ distribution; if it is high entropy then eigenvalue readout can scale exponentially.

Finally, we consider pure states of the form $\rho = \dya{\psi}$. For such states, eigenvalue readout (C3) is efficient because $N_{\text{readout}}$ can be chosen to be independent of $n$, as we noted earlier for the example of pure product states. 

Next we argue that the gate complexity of the diagonalizing unitary, (C1), is efficient. The argument is simply that VQSD takes the state $\rho$ as its input, and $\rho$ must have been prepared on a quantum computer. Let $V$ be the unitary that was used to prepare $\ket{\psi} = V \ket{\vec{0}}$ on the quantum computer. For large $n$, $V$ must have been efficient to implement, otherwise the state $\ket{\psi}$ could not have been prepared. Note that $V\ad$, which is constructed from $V$ by reversing the order of the gates and adjointing each gate, can be used to diagonalize $\rho$. Because $V$ is efficiently implementable, then $V\ad$ is also efficiently implementable. Hence, $\rho$ can be efficiently diagonalized. A subtlety is that one must compile $V\ad$ into one's ansatz, such as the ansatz in Fig.~\ref{fig:iterative-ansatz}. Fortunately, the overhead needed to compile $V\ad$ into our ansatz grows (at worst) only linearly in $n$. An explicit construction for compiling $V\ad$ into our ansatz is as follows. Any one-qubit gate directly translates without overhead into our ansatz, while any two-qubit gate can be compiled using a linear number of swap gates to make the qubits of interest to be nearest neighbors, then performing the desired two-qubit gate, and finally using a linear number of swap gates to move the qubits back to their original positions.

Let us now consider the complexity (C2) of searching for $U$. Since there are a linear number of parameters in each layer, and $p$ needs only to grow polynomially in $n$, then the total number of parameters grows only polynomially in $n$. But this does not guarantee that we can efficiently minimize the cost function, since the landscape is non-convex. In general, search complexity for problems such as this remains an open problem. Hence, we cannot make a general statement about (C2) for pure states.

\section{Implementation of qPCA}\label{sec:appqpca}

In the main text we compared VQSD to the qPCA algorithm. Here we give further details on our implementation of qPCA. Let us first give an overview of qPCA.

\subsection{Overview of qPCA}

\begin{figure}[t]
    \centering
    \includegraphics[scale=0.43]{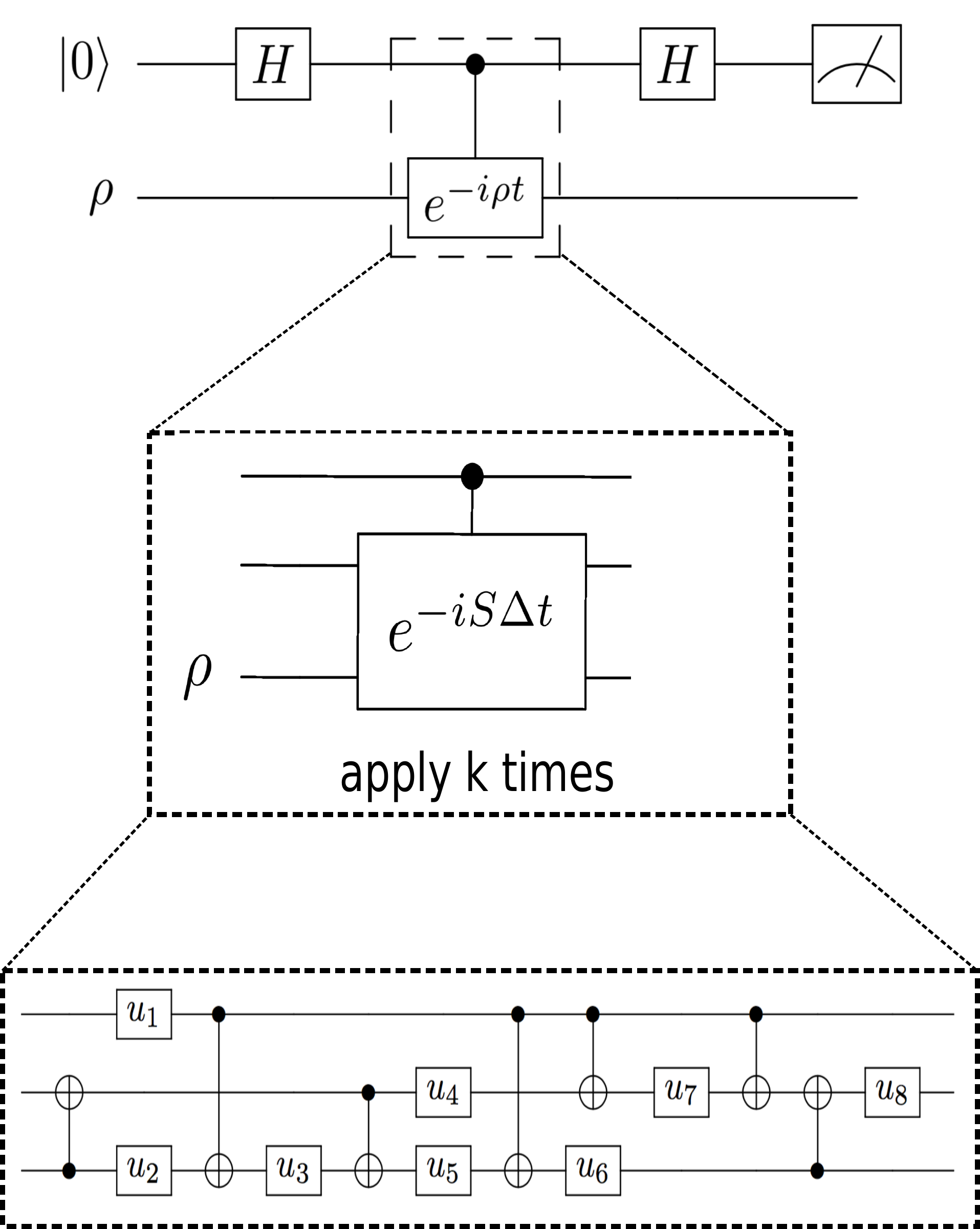}
    \caption{Circuit for our qPCA implementation. Here, the eigenvalues of a one-qubit pure state $\rho$ are estimated to a single digit of precision. We use $k$ copies of $\rho$ to approximate $C_{V(t)}$ by applying the controlled-exponential-swap operator $k$ times for a time period $\Delta t= t/k$. The bottom panel shows our compilation of the controlled-exponential-swap gate into one- and two-qubit gates.}
    \label{fig:qpca-circuit}
\end{figure}

The qPCA algorithm exploits two primitives: quantum phase estimation and density matrix exponentiation. Combining these two primitives allows one to estimate the eigenvalues and prepare the eigenvectors of a state $\rho$.

Density matrix exponentiation refers to generating the unitary $V(t) = e^{-i \rho t}$ for a given state $\rho$ and arbitrary time $t$.
For qPCA, one actually needs to apply the controlled-$V(t)$ gate ($C_{V(t)}$). Namely, in qPCA, the $C_{V(t)}$ gate must be applied for a set of times, $\{t, 2t, 2^2t,...,2^x t\}$, as part of the phase-estimation algorithm. Here we define $t_{\max}:=2^x t$.

Ref.~\cite{lloyd_quantum_2014-1} noted that $V(t)$ can be approximated with a sequence of $k$ exponential swap operations between a target state $\sigma$ and $k$ copies of $\rho$. That is, let $S_{JK}$ be the swap operator between systems $J$ and $K$, and let $\sigma$ and $\rho^{\otimes k}$ be states on systems $A$ and $B = B_1 ... B_r$, respectively. Then one performs the transformation
\begin{align}
    \tau_{AB}= \sigma \otimes (\rho^{\otimes k}) \quad \rightarrow \quad\tau'_{AB} = W(\sigma \otimes (\rho^{\otimes k})) W\ad\,,
\end{align}
where
\begin{align}
\label{eqn:W}
    W= U_{AB_k}\cdots U_{AB_1},\quad\text{and}\quad U_{JK}= e^{-iS_{JK}\Delta t}\,.
\end{align}
The resulting reduced state is 
\begin{align}
\label{eqn:approxV}
    \tau'_A = \Tr_{B}(\tau'_{AB})\approx V(t)\rho V(t)\ad
\end{align}
where $t = k \Delta t$. Finally, by turning each $U_{JK}$ in \eqref{eqn:W} into a controlled operation:
\begin{align}
\label{eqn:controlledES}
    C_{U_{JK}}= \dya{0}\otimes \id + \dya{1}\otimes e^{-iS_{JK}\Delta t}\,,
\end{align}
and hence making $W$ controlled, one can then construct an approximation of $C_{V(t)}$.

If one chooses the input state for quantum phase estimation to be $\rho = \sum_{\vec{z}} \lambda_{\vec{z}} \dya{v_{\vec{z}}}$ itself, then the final state becomes  
\begin{align}
\label{eqn:finalstateqpca}
    \sum_{\vec{z}} \lambda_{\vec{z}} \dya{v_{\vec{z}}} \otimes \dya{\hat{\lambda}_{\vec{z}}}
\end{align}
where $\hat{\lambda}_{\vec{z}}$ is a binary representation of an estimate of the corresponding eigenvalue $\lambda_{\vec{z}}$. One can then sample from the state in \eqref{eqn:finalstateqpca} to characterize the eigenvalues and eigenvectors.

The approximation of $V(t)$ in \eqref{eqn:approxV} can be done with accuracy $\epsilon$ provided that one uses $O(t^2\epsilon^{-1})$ copies of $\rho$. The time $t_{\max}$ needed for quantum phase estimation to achieve accuracy $\epsilon$ is $t_{\max} = O(\epsilon^{-1})$. Hence, with qPCA, the eigenvalues and eigenvectors can be obtained with accuracy $\epsilon$ provided that one uses $O(\epsilon^{-3})$ copies of $\rho$.

\subsection{Our Implementation of qPCA}

\begin{figure}[t!]
    \centering
    \includegraphics[width=\columnwidth]{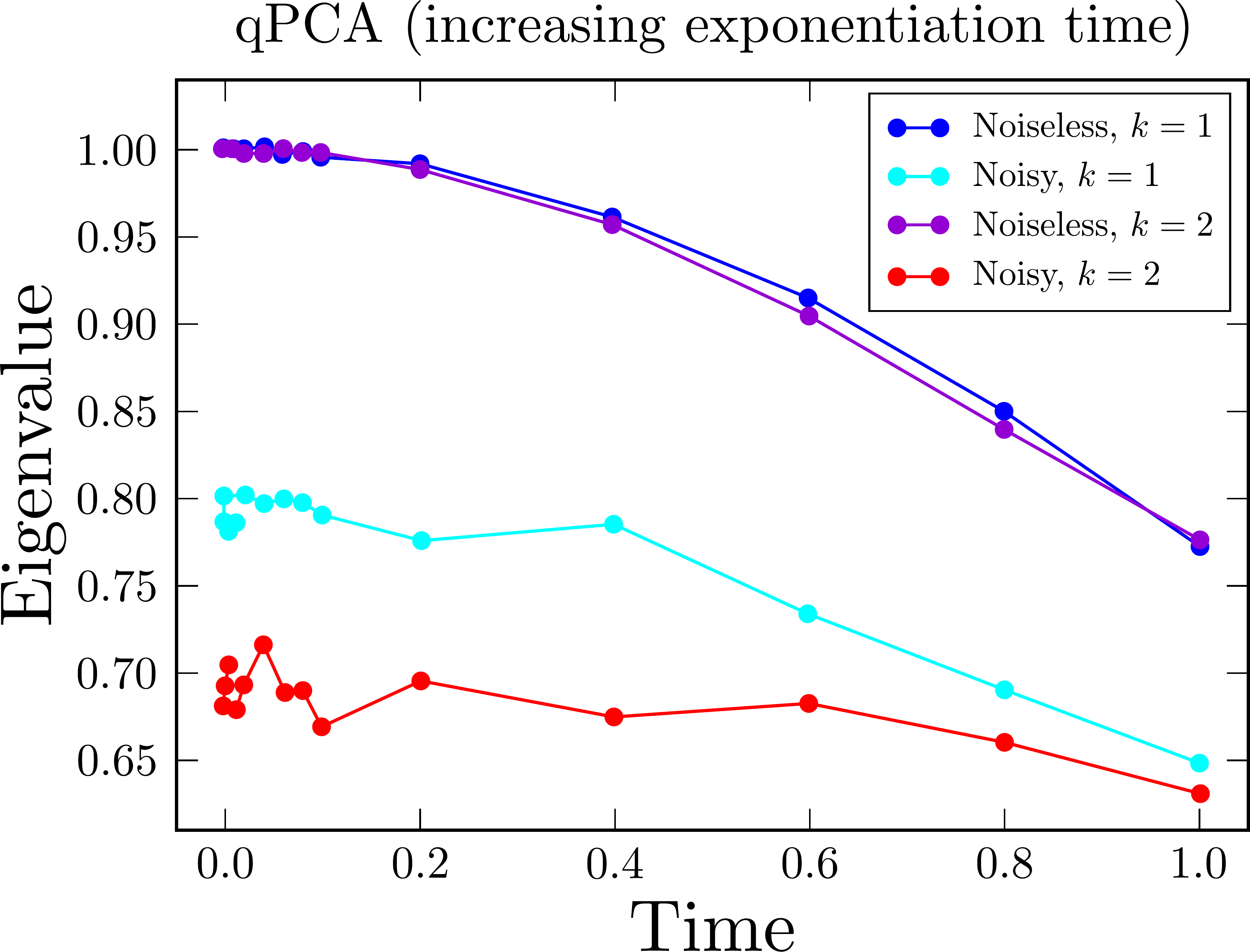}
    \caption{The largest inferred eigenvalue for the one-qubit pure state $\rho=\dyad{+}{+}$ versus application time of unitary $e^{-i\rho t }$, for our implementation of qPCA on Rigetti's noisy and noiseless QVMs. Curves are shown for $k=1$ and $k=2$, where $k$ indicates the number of controlled-exponential-swap operators applied.}
    \label{fig:qPCA_plot}
\end{figure}

Figure~\ref{fig:qpca-circuit} shows our strategy for implementing qPCA on an arbitary one-qubit state $\rho$. The circuit shown corresponds to the quantum phase estimation algorithm with one bit of precision (i.e., one ancilla qubit). A Hadamard gate is applied to the ancilla qubit, which then acts as the control system for the $C_{V(t)}$ gate, and finally the ancilla is measured in the $x$-basis. The $C_{V(t)}$ is approximated (as discussed above) with $k$ applications of the controlled-exponential-swap gate.

To implement qPCA, the controlled-exponential-swap gate in \eqref{eqn:controlledES} must be compiled into one- and two-body gates. For this purpose, we used the machine-learning approach from Ref.~\cite{cincio2018learning} to obtain a short-depth gate sequence for controlled-exponential-swap. The gate sequence we obtained is shown in Fig.~\ref{fig:qpca-circuit} and involves 7 CNOTs and 8 one-qubit gates. Most of the one-qubit gates are $z$-rotations and hence are error-free (implemented via a clock change), including the following gates:
\begin{align}
    &u_{1}= u_{7} = R_{z}(-(\pi+\Delta t)/2)\\ 
    &u_{3}= R_{z}((\pi-\Delta t)/2)\\
    &u_{4}= R_{z}(\Delta t/2)\\
    &u_{5}= R_{z}((\pi+\Delta t)/2) \\
    &u_{8}= R_{z}(\pi/2)\,.
\end{align}
The one-qubit gates that are not $z$-rotations are:
\begin{align}
&u_2 = \frac{1}{\sqrt{2}}
\begin{pmatrix}
   1 & 1 \\
   e^{-i(\pi-\Delta t)/2} & e^{i(\pi+\Delta t)/2} \\
\end{pmatrix}\\
&u_6 = \frac{1}{\sqrt{2}}
\begin{pmatrix}
   1 & e^{-i(\pi+ \Delta t)/2} \\
   -i & e^{-i \Delta t/2}
\end{pmatrix}\,.
\end{align}

We implemented the circuit in Fig.~\ref{fig:qpca-circuit} using both Rigetti's noiseless simulator, known as the Quantum Virtual Machine (QVM), as well as their noisy QVM that utilizes a noise model of their 8Q-Agave chip. Because the latter is meant to mimic the noise in the 8Q-Agave chip, our qPCA results on the noisy QVM can be compared to our VQSD results on the 8Q-Agave chip in Fig.~\ref{fig:oneq-qpu-data}. (We remark that lack of availability prevented us from implementing qPCA on the actual 8Q-Agave chip.)

For our implementation, we chose the one-qubit plus state, $\rho = \dya{+}$. Implementations were carried out using both one and two controlled-exponential-swap gates, corresponding to $k=1$ and $k=2$. The time $t$ for which the unitary $e^{-i\rho t }$ was applied was increased.

Figure~\ref{fig:qPCA_plot} shows the raw data, i.e., the largest inferred eigenvalue versus $t$. In each case, small values of $t$ gave more accurate eigenvalues. In the noiseless case, the eigenvalues of $\rho=\dyad{+}{+}$ were correctly estimated to be $\approx \{1,0\}$ already for $k=1$ and consequently also for $k=2$. In the noisy case, the eigenvalues were estimated to be $\approx \{0.8,0.2\}$ for $k=1$ and $\approx \{0.7,0.3\}$ for $k=2$, where we have taken the values for small $t$. Table~\ref{table:2} summarizes the different cases.

\begin{table}[t!]
\begin{tabular}{|c|c|c|} 
 \hline
 QVM & $k=1$ & $k=2$ \\ 
 \hline
 noiseless & $\approx \{1,0\}$ & $\approx \{1,0\}$  \\ 
 \hline
 noisy & $\approx \{0.8, 0.2\}$ & $\approx \{0.7,0.3\}$  \\
 \hline
 \end{tabular}
\caption{Estimated eigenvalues for the $\rho=\dyad{+}{+}$ state using qPCA on both the noiseless and the noisy QVMs of Rigetti. }
\label{table:2}
\end{table}

Already for the case of $k=1$, the required resources of qPCA (3 qubits + 7 CNOT gates) for estimating the eigenvalue of an arbitary pure one-qubit state $\rho$ are higher than those of the DIP test (2 qubits + 1 CNOT gate) for the same task. Consequently, the DIP test yields more accurate results as can be observed by comparing Fig. \ref{fig:oneq-qpu-data} to Fig. \ref{fig:qPCA_plot}. Increasing the number of copies to $k=2$ only decreases the accuracy of the estimation, since the $C_{V(t)}$ gate is already well approximated for short application times $t$ when $k=1$ in the noiseless case. Thus, increasing the number of copies does not offer any improvement in the noiseless case, but instead leads to poorer estimation performance in the noisy case. This can be seen for the $k=2$ case (see Fig. \ref{fig:qPCA_plot} and Table \ref{table:2}), due to the doubled number of required CNOT gates relative to $k=1$.

\section{Circuit derivation} \label{app:cd}

\subsection{DIP test}

Here we prove that the circuit in Fig.~\ref{fig:supp_DIP}(a) computes $\Tr(\ZC(\sigma) \ZC(\tau))$ for any two density matrices $\sigma$ and $\tau$.

Let $\sigma$ and $\tau$ be states on the $n$-qubit systems $A$ and $B$, respectively. Let $\omega_{AB} = \sigma \otimes \tau$ denote the initial state. The action of the CNOTs in Fig.~\ref{fig:supp_DIP}(a) gives the state
\begin{align}
 \omega'_{AB} & =   \sum_{\vec{z}, \vec{z'}} X^{\vec{z}}\sigma X^{\vec{z'}} \otimes \dya{\vec{z}}\tau\dya{\vec{z'}},
\end{align}
where the notation $X^{\vec{z}}$ means
$X^{z_1}\otimes X^{z_2} \otimes \cdots \otimes X^{z_n}$. Partially tracing over the $B$ system gives
\begin{align}
 \omega'_{A} & =   \sum_{\vec{z}} \tau_{\vec{z},\vec{z}} X^{\vec{z}}\sigma X^{\vec{z}}\,,
\end{align}
where $\tau_{\vec{z},\vec{z}} = \bra{\vec{z}}\tau\ket{\vec{z}}$. The probability for the all-zeros outcome is then
\begin{align}
 \bra{\vec{0}}\omega'_{A}\ket{\vec{0}} & =   \sum_{\vec{z}} \tau_{\vec{z},\vec{z}} \bra{\vec{0}}X^{\vec{z}}\sigma X^{\vec{z}}\ket{\vec{0}}  =  \sum_{\vec{z}} \tau_{\vec{z},\vec{z}}  \sigma_{\vec{z},\vec{z}}, 
\end{align}
which follows because $X^{\vec{z}}\ket{\vec{0}} = \ket{\vec{z}}$. Hence the probability for the all-zeros outcome is precisely the diagonalized inner product, $\Tr(\ZC(\sigma) \ZC(\tau))$. Note that in the special case where $\sigma = \tau = \rhot$, we obtain the sum of the squares of the diagonal elements, $\sum_{\vec{z}}  \rhot_{\vec{z},\vec{z}}^2 = \Tr(\ZC(\rhot)^2)$.

\subsection{PDIP test}

\begin{figure}
    \centering
    \includegraphics[width = \linewidth]{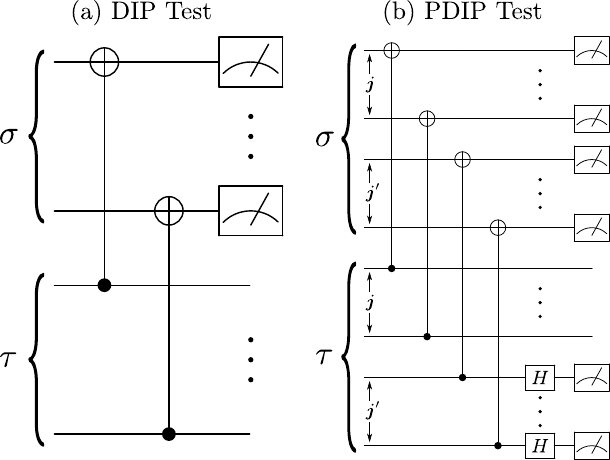}
    \caption{Test circuits used to compute the cost function in VQSD. (a) DIP test (b) PDIP test. (These circuits appear in Fig.~\ref{fig:diptest} and are also shown here for the reader's convenience.)
   }
    \label{fig:supp_DIP}
\end{figure}

We prove that the circuit in Fig.~\ref{fig:supp_DIP}(b) computes $\Tr(\ZC_{\vec{j}}(\sigma)\ZC_{\vec{j}}(\tau))$ for a given set of qubits $\vec{j}$.

Let $\vec{j}'$ denote the complement of $\vec{j}$. Let $\sigma$ and $\tau$, respectively, be states on the $n$-qubit systems $A = A_{\vec{j}\vec{j}'}$ and $B = B_{\vec{j}\vec{j}'}$. The initial state $\omega_{AB} = \sigma \otimes \tau$ evolves, under the action of the CNOTs associated with the DIP Test and then tracing over the control systems, to
\begin{align}\label{eqn:cost-definition_alt5}
 \omega'_{AB_{\vec{j}'}} & = \sum_{\vec{z}} (X^{\vec{z}}\otimes \id)\sigma (X^{\vec{z}}\otimes \id) \otimes \Tr_{B_{\vec{j}}}((\dya{\vec{z}}\otimes \id)\tau),
\end{align}
where $X^{\vec{z}}$ and $\dya{\vec{z}}$ act non-trivially only on the $\vec{j}$ subsystems of $A$ and $B$, respectively. Measuring system $A_{\vec{j}}$ and obtaining the all-zeros outcome would leave systems $A_{\vec{j}'}B_{\vec{j}'}$ in the (unnormalized) conditional state:
\begin{align}\label{eqn:cost-definition_alt6}
 \Tr_{A_{\vec{j}}}((\dya{\vec{0}}\otimes \id)\omega'_{AB_{\vec{j}'}}) & = \sum_{\vec{z}}\sigma_{\vec{j}'}^{\vec{z}} \otimes \tau_{\vec{j}'}^{\vec{z}}, 
\end{align}
where $\sigma_{\vec{j}'}^{\vec{z}}:=\Tr_{A_{\vec{j}}}((\dya{\vec{z}}\otimes \id)\sigma)$ and $\tau_{\vec{j}'}^{\vec{z}} :=\Tr_{B_{\vec{j}}}((\dya{\vec{z}}\otimes \id)\tau)$. Finally, computing the expectation value for the swap operator (via the Destructive Swap Test) on the state in \eqref{eqn:cost-definition_alt6} gives
\begin{align}\label{eqn:cost-definition_alt7}
  \sum_{\vec{z}} \Tr((\sigma_{\vec{j}'}^{\vec{z}} \otimes \tau_{\vec{j}'}^{\vec{z}})S) = \sum_{\vec{z}} \Tr(\sigma_{\vec{j}'}^{\vec{z}} \tau_{\vec{j}'}^{\vec{z}}) = \Tr(\ZC_{\vec{j}}(\sigma)\ZC_{\vec{j}}(\tau))\,.
\end{align}
The last equality can be verified by noting that $\ZC_{\vec{j}}(\sigma) = \sum_{\vec{z}} \dya{\vec{z}} \otimes \sigma_{\vec{j}'}^{\vec{z}}$ and $\ZC_{\vec{j}}(\tau) = \sum_{\vec{z}} \dya{\vec{z}} \otimes \tau_{\vec{j}'}^{\vec{z}}$. Specializing \eqref{eqn:cost-definition_alt7} to $\sigma = \tau = \rhot$ gives the quantity $\Tr(\ZC_{\vec{j}}(\rhot)^2)$.

\section{Proof of \texorpdfstring{Eq.~\eqref{eqn:c2bound}}{Eq.}} \label{sec:appProof}

Let $H_2(\sigma) = - \log_2 [ \Tr (\sigma^2)]$ be the Renyi entropy of order two. Then, noting that $H_2(\sigma) \leq H(\sigma)$, we have 
\begin{align}\label{eqn:proof1}
\Tr(\ZC_j(\rhot)^2) = 2^{-H_2(\ZC_j(\rhot))}\geq 2^{-H(\ZC_j(\rhot))}\,.
\end{align}
Next, let $A$ denote qubit $j$, and let $B$ denote all the other qubits. This allows us to write $\rho = \rho_{AB}$ and $\rhot = \rhot_{AB}$. Let $C$ be a purifying system such that $\rho_{ABC}$ and $\rhot_{ABC}$ are both pure states. Then we have
\begin{align}
\label{eqn:proof2}
H(\ZC_j(\rhot)) &= H(\ZC_j(\rhot_{AB})) \\
\label{eqn:proof3}
&= H(\ZC_j(\rhot_{AC})) \\
\label{eqn:proof4}
&\leq H(\ZC_j(\rhot_{A})) +H(\rhot_{C})
\end{align}
where the inequality in \eqref{eqn:proof4} used the subadditivity of von Neumann entropy. Finally, note that \begin{align}
\label{eqn:proof5}
H(\rhot_{C}) = H(\rhot_{AB} ) = H(\rho_{AB} ) = H(\rho)
\end{align}
and $H(\ZC_j(\rhot_{A})) \leq 1$, which gives
\begin{align}
\label{eqn:proof6}
H(\ZC_j(\rhot))\leq 1 + H(\rho)\,.
\end{align}
Substituting \eqref{eqn:proof6} into \eqref{eqn:proof1} gives
\begin{align}
\label{eqn:proof7}
\Tr(\ZC_j(\rhot)^2) \geq 2^{-1 - H(\rho)}\,,
\end{align}
and \eqref{eqn:c2bound} follows from $H(\rho)\leq \log_2 r$.

%% file: vqsd.bbl
\begin{thebibliography}{52}
\bibitem [{foo()}]{footnote1}%
  \BibitemOpen
  \href@noop {} {}\bibinfo {note} {This can be seen by noting that computing
  eigenvalues is equivalent to computing roots of a polynomial equation (namely
  the characteristic polynomial of the matrix) and that no closed-form solution
  exists for the roots of general polynomials of degree greater than or equal
  to five \cite{trefethen97}.}\BibitemShut {Stop}%
\end{thebibliography}
